# Data Poisoning Attacks in Intelligent Transportation Systems: A Survey

Feilong Wang[a], Xin Wang[b], Xuegang (Jeff) Ban[b,*]

[a]School of Transportation and Logistics, Southwest Jiaotong University, Chengdu, 610032, China

[b]Department of Civil and Environmental Engineering, University of Washington, Seattle, WA, 98105, United States

[*]Corresponding author: banx@uw.edu

**Abstract**

Emerging technologies drive the ongoing transformation of Intelligent Transportation Systems (ITS). This transformation has given rise to cybersecurity concerns, among which data poisoning attack emerges as a new threat as ITS increasingly relies on data. In data poisoning attacks, attackers inject malicious perturbations into datasets, potentially leading to inaccurate results in offline learning and real-time decision-making processes. This paper concentrates on data poisoning attack models against ITS. We identify the main ITS data sources vulnerable to poisoning attacks and application scenarios that enable staging such attacks. A general framework is developed following rigorous study process from cybersecurity but also considering specific ITS application needs. Data poisoning attacks against ITS are reviewed and categorized following the framework. We then discuss the current limitations of these attack models and the future research directions. Our work can serve as a guideline to better understand the threat of data poisoning attacks against ITS applications, while also giving a perspective on the future development of trustworthy ITS.

## 1. Introduction

The ongoing transformation of Intelligent Transportation Systems (ITS) driven by emerging technologies like mobile sensing, new mobility systems, advanced driving systems, transportation electrification, and the widespread application of artificial intelligence (AI), is reshaping human mobility, urban landscapes, and society at large (Ban and Gruteser, 2012; Herrera et al., 2010; Wang et al., 2019; Wang and Chen, 2018). This transformation has given rise to emerging cybersecurity concerns as ITS increasingly relies on these technologies, making them more susceptible to cyber threats and attacks. Researchers have responded to this issue through extensive research, focusing on automotive cybersecurity, in-vehicle networks (such as the Controller Area Network bus, electronic control units, and onboard sensors), and vehicle-to-everything (V2X) communications, particularly in the context of connected and automated vehicles (CAVs) (Cui et al., 2019; Petit and Shladover, 2015).

One cybersecurity concern in transportation that is gaining increasing attention comes from the rapid accumulation of diverse datasets, spanning vehicle and user-generated data, infrastructure information, communication between vehicles and infrastructure, and data from sources like social media. These datasets are revolutionizing various aspects of transportation, including traffic prediction, control, safety, and human mobility analysis. A new threat emerges from the data-driven paradigm, the so-called "Data Poisoning Attacks (DPAs)", where adversaries inject malicious perturbations into a dataset, potentially leading to inaccurate results if the data is used for offline training and/or real-time decision-making processes (Cinà et al., 2023; Wang et al., 2024a). This could cause severe system-level impacts, e.g., (sizable) errors in volume or queue estimation at signalized intersections could results in improper signal timing plans and subsequently more delays. Errors in travel time estimation or congestion identification (classification) could mislead vehicles to select congested routes (misclassed as free flowing or lightly congested), leading to more congestion or even gridlocks. While data attacks like poisoning are well-studied in cybersecurity, their application in transportation, encompassing various data-related attacks, presents a unique set of challenges that need to be rigorously studied. Note that terms such as evasion attacks (Jiang et al., 2020) or false data injection attacks (Zhao et al., 2022a) were also used for data attacks related to real-time decisions, while data poisoning often refers to attacks on the training data. Here, with a little abuse of the terminology, we use *data poisoning* (or DPA in short) as a general term to denote a broad spectrum of attacks that compromise ITS by manipulating input data, including attacks on both the training stage and decision (inference) stage. Note that training and decision attacks are also referred to as offline and online attacks in some literature. As noted below and detailed in Section 3.1, this review focuses on DPAs against primary data sources supporting ITS applications.

Recent studies show that DPAs are of high concern for deploying emerging ITS technologies due to their high likelihood and diverse motivations (Wang et al., 2024a). Attackers target ITS for a multitude of reasons, including disrupting traffic flow (Feng et al., 2022), stealing sensitive data (Sun et al., 2021a), financial gain through extortion or exploiting tolling systems (Wang et al., 2023a), cyberterrorism aimed at instilling fear or achieving political objectives (Tegler, 2023), and sabotaging infrastructure to undermine economic and social stability (Shirvani et al., 2024). Several sources confirm that DPAs against ITS have already been conducted in practice to reveal their potential damages. For example, Global Navigation Satellite System (GNSS) spoofing, which broadcasts falsified GNSS signals, has been a long-recognized high threat (Schmidt et al., 2016; Tegler, 2023); Light Detection and Ranging sensor (LiDAR) can be compromised by replay attacks that deceive receivers with recorded and thus outdated data (Cao et al., 2019c); cameras are sensitive to blinding attacks that emit light into the camera (Petit et al., 2015; Petit and Shladover, 2015). A recent study has demonstrated that a data-fusion-based navigation system can be poisoned, disappointing system designers who desire to ensure the robustness of systems via redundant sensors (Shen et al., 2020). Data shared among vehicles (e.g., basic safety message (BSM)) can be maliciously falsified by breaking into the V2X communication channels and modifying the transmitted data (Boeira et al., 2018). In vehicle platooning, the falsified BSM could break the platooning stability (Dadras et al., 2015). Besides the decision-time attacks, studies have shown that vehicle sensors and systems can also be attacked during training times. Jiang et al. (2020) have shown that the deep-learning (DL)-based computer vision system deployed on a vehicle can be attacked by feeding adversarial traffic signs into the training dataset. At the network level, attacks can also be launched to create false jams with ghost vehicles on the roads, affecting the path-routing algorithms and decisions of massive travelers (Yu, 2021). Recent studies also show that DL-based network-level traffic prediction models can be manipulated by poisoning a small set of traffic sensors, reducing network efficiency.

Given these concerns, a systematic survey of threats from DPAs in ITS is imperative to understand the state-of-the-art research in this domain, and to identify research gaps and future research needs. Existing surveys in the literature on transportation cybersecurity are specific to an application domain (e.g., onboard sensors, CAVs, or V2X) or only consider a high-level overview of the whole spectrum of attacks in transportation (e.g., on the hardware- or software-related vulnerability in ITS systems) (Abdo et al., 2023; Almalki and Song, 2020; Alnasser et al., 2019; Cui et al., 2019; Deng et al., 2021; Ju et al., 2022; Kuutti et al., 2018; Marcillo et al., 2022; Salek et al., 2022). A few review papers on DPAs are on computer vision systems (i.e., data in image or video formats), which only covers a specific data source for transportation applications (Biggio and Roli, 2018; Cinà et al., 2023; Goldblum et al., 2023; Ramirez et al., 2022; Tian et al., 2022). In this survey paper, we concentrate on and review current DPA models against ITS data. We identify the primary ITS data sources vulnerable to DPAs and application scenarios that enable staging such attacks. Attacks against data for three primary ITS components are systematically reviewed and categorized, including individual road users (vehicles and other users), V2X communication, and the infrastructure. We then propose a formal study framework that follows rigorous study process from cybersecurity but also considers critical ITS application needs such as risk assessment and system impact of the attacks. We further discuss the current main limitations of attack models and the future research directions.

Our work can serve as a guideline to understand the threat of DPAs to ITS systems (e.g., how and when these attacks can be staged). The understanding is crucial for several reasons. Firstly, it helps to understand the tactics and techniques employed by adversaries to manipulate ITS data, thus enhancing awareness of potential vulnerabilities in data-driven ITS applications. Secondly, it facilitates developing effective defense solutions and countermeasures to mitigate the impact of such attacks. Thirdly, studying DPA models aids in evaluating the robustness of ITS systems, providing insights on future design of trustworthy ITS. All these help enhance the overall cybersecurity of ITS and the entire transportation system.

Figure 1 provides a diagram of how we plan to conduct the review, which also indicates the organization of the paper. In Section 2, we start with an overview of the general framework to review and synthesize DPA models by adopting the state-of-the-art data poisoning attack framework in cybersecurity, enhanced by unique features and considerations for ITS applications. Section 3 reviews existing DPAs in ITS. It first provides a general background of ITS data from three primary sources to help readers understand the vulnerabilities of ITS to DPAs, including data supporting individual users, data from V2X communications, and data from the infrastructure. Then, DPAs against each of the three data sources in ITS are reviewed. Section 4 synthesizes the reviews in Section 3 by following the general framework, focusing on discussing major components of DPAs and obtaining insights for risk assessment. Section 5 and 6 discuss the limitations of current attack models and the future research directions, respectively. Section 7 concludes the work.

To summarize, this work provides the following major contributions:

i. We present the first review, to the authors' best knowledge, that focuses on DPAs against the primary data sources in ITS.
ii. We categorize existing DPA models in ITS according to the data source they are targeting and the functions the data supports.
iii. We provide a general framework to model DPAs in ITS while accounting for unique ITS application needs such as risk assessment.
iv. We discuss limitations of existing attack models, open questions, and challenges regarding DPAs against ITS.

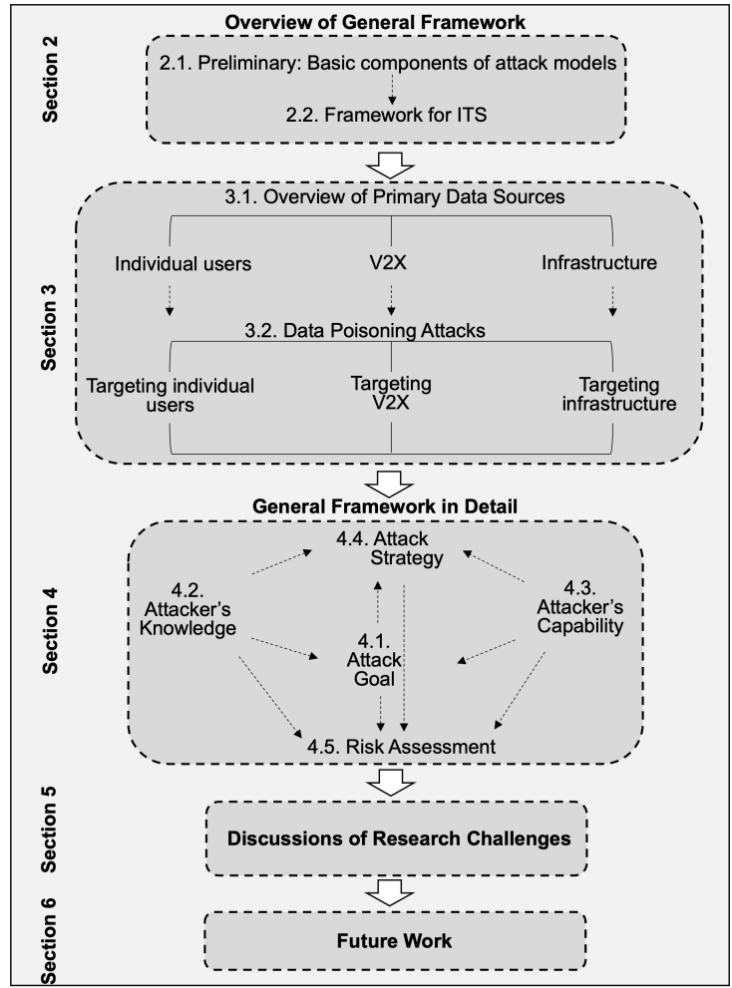

Figure 1. Diagram of the review

## 2. An Overview of the General Framework for DPAs in ITS

2.1. Preliminary: Basic Components of Attack Models

A DPA aims to cause harm by systematically manipulating the data used in a data-driven model. Formal study of such attacks plays a pivotal role in understanding and defending against these malicious endeavors. A DPA typically consists of four major components, including the attack goal, attacker's knowledge, attacker's capability and attack strategy (Figure 6; Reda et al., 2021). Each component is briefly introduced below; a simple example can be found in Appendix A.1 to illustrate each component using a Support Vector Machine (SVM) model for vehicle classification.

The *attack goal* defines what the attacker wants to achieve through the attack. The attack goals of DPAs in the cybersecurity domain mainly focus on how the attack influences the model's classification, predictions, or other data-oriented tasks. For instance, the attack could be error-specific, where the attacker seeks to have a sample misclassified as a specific class (Melis et al., 2017), or error-generic, where the attacker aims to have a sample misclassified as any class different from the true class (Moosavi-Dezfooli et al., 2016).

The *attacker's knowledge* pertains to how the attacker understands the target system and its components, encompassing the model (parameters) and input data. It may include the knowledge of a path by which the attacker can access and manipulate a victim's system physically or remotely. Depending on attacker's knowledge, there could be different types of attack settings, including white-box and black-box attacks, with grey-box attack in the middle. In the *white-box attack* setting, the attacker possesses complete knowledge about the target system. Although not always representative of real-world scenarios, white-box attacks allow for a worst-case analysis and are valuable for evaluating defense mechanisms (Biggio and Roli, 2018). In contrast, in the *black-box attack s*etting, the attacker has no (or little) knowledge of the training data and/or the target model. Black-box attacks can be transfer attacks or query attacks. The former utilizes a surrogate model (e.g., a public model) to approximate the target and then applies white-box attacks against the surrogate model (Melis et al., 2017). The latter relies on input queries and model predictions to iteratively refine the attack, without additional knowledge (Bhagoji et al., 2018).

*Attacker's capability* specifies the resources an attacker needs to launch a successful attack. The attacker could have chances to manipulate a specific *learning settings/process*, such as training-from-scratch, fine-tuning, and model-training process. The attacker may alter a subset of the training dataset, manipulate the training process, or act as a man-in-the-middle to control the training process. Attacker's capability also influences the way how data is perturbed (Goldblum et al., 2023).

*Attack strategy* details how the attacker manipulates/perturbs data to execute the desired poisoning attack and achieve the attack goal. A sophisticated algorithm is often needed to ensure a stealthy (i.e., inconspicuous perturbations to data) and effective attack. Typically, a set of optimal perturbations is determined through *optimization problems*, such as bilevel programming, which identify perturbations for achieve attack goal while satisfying specific constraints (Fang et al., 2020).

2.2. Framework for Studying DPAs in ITS

We propose a comprehensive framework to study DPAs in ITS that reviews, categorizes, assesses, and synthesizes existing DPAs within ITS from a holistic perspective. This section provides an overview of the framework for ITS. More elaborations are provided in Section 4, following our review of existing studies on DPAs in ITS.

As shown in Figure 2, the attack framework adheres to the established process in studying DPAs in the cybersecurity domain, around the four major components: attack goal, attacker's knowledge, attacker's capability, and attack strategy. The precise structure within each component is tailored to the unique characteristics of ITS. Interactions can be observed among the four components defining an attack model in ITS. When setting a goal, an attacker has in mind his/her knowledge of the target and capability of conducting the attack. As a result, the attack goal directs the attacker's actions in data manipulation, which are confined by attacker's knowledge and capability. Moreover, how to execute the desired poisoning attack relies on attack goal, knowledge, and capability. Therefore, these three components collectively shape the fourth component, the attack strategy.

For ITS cybersecurity, risk assessment is an essential component that assesses the risk of potential attacks, as emphasized in the Transportation System Section Cybersecurity Framework Implementation Guidance by U.S. Department of Homeland Security (2015). As illustrated in Figure 2, risk assessment in this context is typically computed as the product of impact and likelihood (National Academies of Sciences, 2020; Feng et al., 2022). Different from the direct influence measured by attack goal, the impact may represent an attack's system-level impact. The likelihood is affected by the easiness and stealthiness of executing the designed attack (Feng et al., 2022) which are relevant to attacker's knowledge, capability, and attack strategy. The product of the two leads to the risk assessment matrix, which is important for enhancing ITS security. For instance, the matrix identifies attacks of both high impact and high likelihood that shall be given the high priority to defense (Figure 2).

Current research on risk assessment on DPAs is sparse and the only study we are aware of is Feng et al. (2022). The proposed attack framework in Figure 2 provides a conceptual process for risk assessment: the various levels of influence in attack goals enable assessing potential attack impacts, while the categories that account for attacker knowledge, capability, and attack strategy provide a basis for evaluating the likelihood of successfully executing these attacks. Section 4.3 provides more discussion on using this conceptual process to conduct some initial risk assessment. More future research in this area is critically needed as detailed in Section 6.

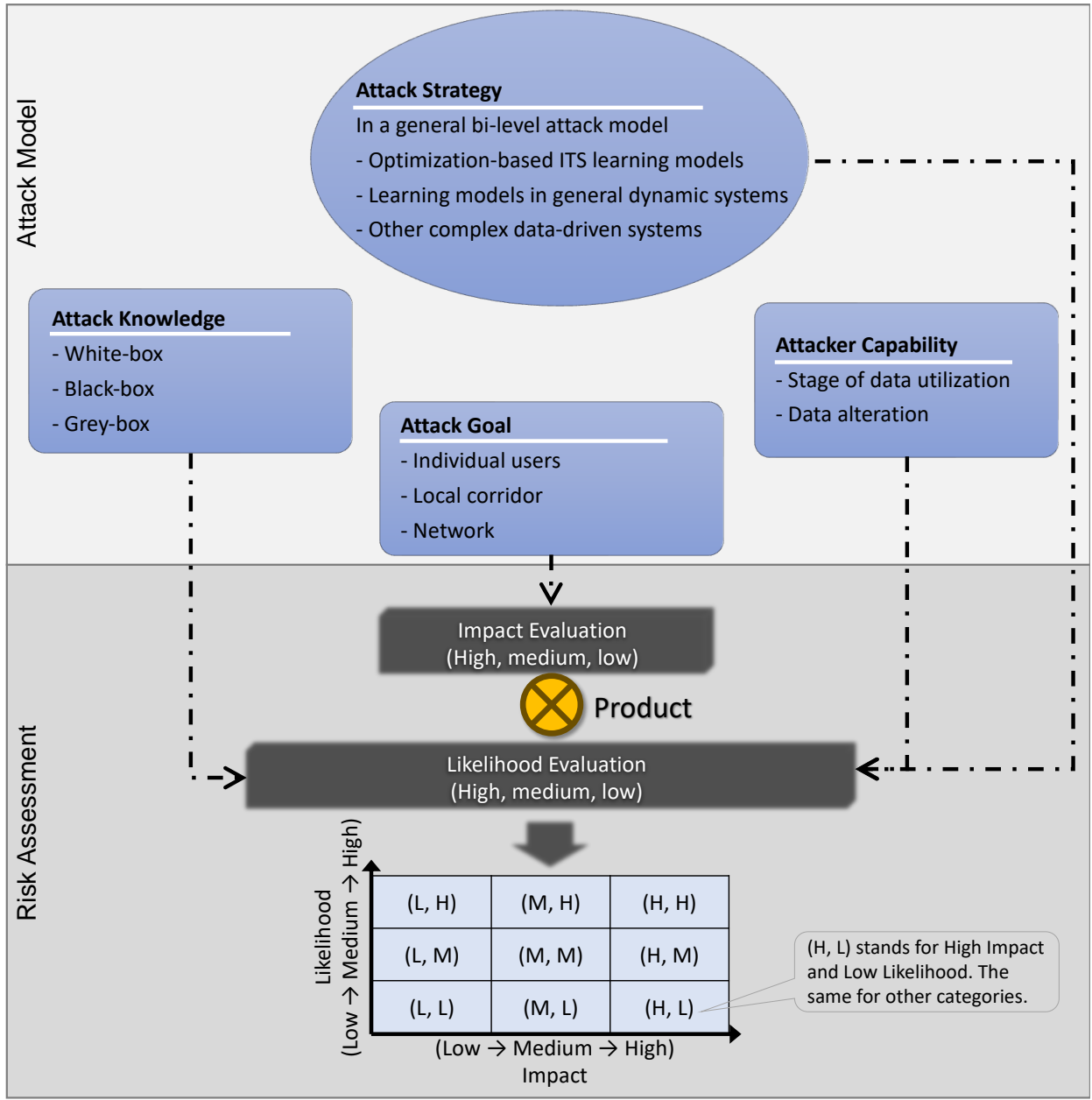

Figure 2. A general framework of DPAs and risk assessment in ITS.

## 3. DPAs against ITS data

In general, ITS consists of three major components: intelligent vehicles/users, intelligent infrastructure (roads), and the communication between vehicles/users and the infrastructure. The transformations in ITS are progressively steering these components to become more data driven. On the *vehicle* side, advanced vehicles are equipped with various sensors and heavily rely on AI models trained on vast datasets to perceive their surroundings and support advanced driving. Simultaneously, on the *infrastructure* side, sensing and communicating technologies (e.g., cameras, Lidar, etc.) are increasingly deployed to support applications such as intelligent traffic signal control (TSC) that relies on trajectories from cameras to enhance traffic safety and efficiency. *V2X communication* is increasingly involved in many aspects of ITS, particularly enabling the interaction between vehicles and other users and between vehicles and the infrastructure. As depicted in Figure 3, data from these three primary data sources within ITS exhibit interconnections, implying that an attack directed at one source may have far-reaching consequences, affecting other data sources indirectly. These interactions suggest that besides the direct impact defined by the attack goal, an attack could have long-reaching impacts, suggesting the necessity of risk assessment. For instance, CAVs can be vulnerable to attacks that compromise the supportive infrastructure, and conversely, the infrastructure's integrity can be undermined by attacking against data from vehicles or V2X.

Each of the three primary data sources is threatened by DPAs. We examine the attacks to each data source below. To assist the understanding of these attacks, we first introduce each data source and its role. We then group the current attack models around the three primary data sources. It is important to note that attacks can also be categorized from different perspectives, such as the learning models under attack, including optimization-based learning models, classical machine learning (ML) models, DL models, reinforcement learning (RL) models, federated learning models, and more (Al Mallah et al., 2021). However, given the diverse range of applications within ITS, only a limited number of DPA cases involve these specific learning models. Many victim systems within ITS take various forms, such as the dynamics of automated vehicles, planning schemes for vehicle platooning, and control algorithms for TSC. Hence, rather than categorizing attacks based on the underlying models, this review is organized by data sources. Section 4 provides more discussions on the underlying learning models DPAs may target.

## 3.1. Overview of Primary ITS Data Sources

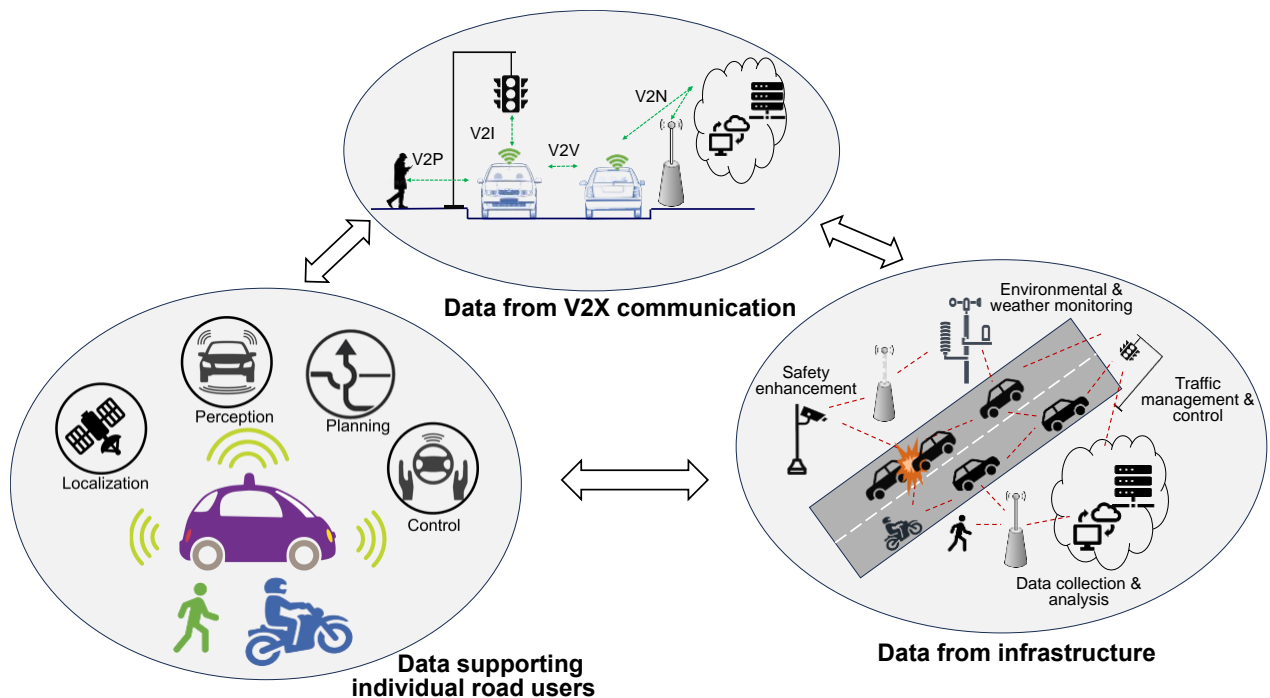

Figure 3. Primary data sources in ITS

### 3.1.1. Data supporting individual users

Road users encompass not only vehicles but also others including pedestrians and bicyclists, whose data may be susceptible to attacks with potentially harmful consequences. Surprisingly, while numerous studies have investigated attacks on vehicular data, there is a lack of research dedicated to understanding the threats of attacks on pedestrian and bicyclist data specifically. Therefore, this survey paper focuses solely on examining attacks on vehicular data, recognizing the need for future research to address the security implications surrounding pedestrian and bike data (see Section 6).

An advanced vehicle integrates multiple technologies, including vehicle localization, environmental perception, path planning, motion control, etc. Accurate *vehicle localization* is critical due to its core role in planning vehicle routes and executing controls (Kuutti et al., 2018). GNSS has been widely used as a positioning input, combined with other sensors (Shen et al., 2020; Wang et al., 2021b). LiDAR-based methods localize a vehicle by scanning the surroundings using lasers and matching the scans with a High-Definition Map in the database (Gao et al., 2015). Similarly, simultaneous localization and mapping (SLAM) is another advanced algorithm that determines the current position of a vehicle based on the observed environmental characteristics that are obtained by scanning the surroundings. The modern localization solutions are predominantly based on data-fusion algorithms that combine position information from multiple sensors, including GNSS, Inertial Measurement Unit (IMU), and LiDAR (Shen et al., 2020), for high-accuracy positioning and robustness under different road and weather conditions (Cao et al., 2019c; Gao et al., 2015). *Vehicle perception* system typically relies on data from several sensors collectively to extract vital information from the vehicle's surroundings (Guerrero-Ibáñez et al., 2018). Ultrasonic radars are employed for low-speed applications like parking assistance due to their relatively low response speed and resolution. Designed for medium to long-range ranging, millimeter-wave radar is resilient in harsh weather conditions (Venon et al., 2022). LiDAR employs laser transmitters and receivers to construct real-time 3D point cloud maps, offering high precision and long-distance object detection (Beltrán et al., 2018). *Cameras* capture images that are analyzed to extract key features for image recognition, mainly responsible for recognizing traffic lights, signs, and other objects (Yazdi and Bouwmans, 2018). *Trajectory planning* involves the computation of a path that guides the vehicle's movements, ensuring it reaches its destination while adhering to traffic regulations, avoiding obstacles, and optimizing efficiency. Diverse data sources, such as vehicle state data, real-time traffic conditions, and environmental variables, are needed to perform trajectory planning effectively (Dixit et al., 2018). Once a CAV has completed the localization, perception and planning, *vehicle control* translates these actions into controlled vehicle operations. Vehicle control typically encompasses two fundamental aspects: lateral control, which involves adjusting the steering wheel and managing lateral tire forces, and longitudinal control, which controls acceleration and braking (Eskandarian et al., 2019). Feedback control mechanisms, notably the proportional–integral–derivative controller, are widely employed for vehicle control.

### 3.1.2. Data from V2X communication

Connectivity transforms vehicles into integral components of the Internet of Vehicles, which relies on data-centric communication mechanisms collectively known as V2X (Ghosal and Conti, 2020). V2X encompass Vehicle-to-Vehicle (V2V), Vehicle-to-Infrastructure (V2I), Vehicle-to-Network (V2N), and Vehicle-to-Pedestrian (V2P) communications. *V2V*, common on urban streets and highways, facilitates peer-to-peer data sharing, enabling vehicles to exchange speed, acceleration, braking, relative positions, and steering information for proactive safety measures. Vehicular Ad Hoc Networks (VANETs), driven by V2V, underpin various ITS advancements, including vehicle platooning, vehicle automated braking, traffic information dissemination, and emergency services (Arif et al., 2019). *V2I* involves vehicle-mounted devices interfacing with Roadside Units (RSUs), fostering real-time data exchange between vehicles and infrastructure. Traffic management centers harness this data for optimizing traffic signals, incident management, and traffic flow enhancement (Nallaperuma et al., 2019). Via V2I, infrastructure can distribute real-time information to travelers via diverse channels, empowering vehicles and other users with real-time traffic data and surrounding information. *V2N* employs robust, high-speed network infrastructure (e.g., 5G Cellular-V2X) for seamless data transmission between vehicles and management systems (Cao et al., 2019a). High-speed communication infrastructures support data exchange across long-distance vehicles, infrastructure, and traffic management centers. Cloud servers store and analyze this data, fostering services like navigation, remote monitoring, emergency assistance, and entertainment (Deng et al., 2020). *V2P* technology aims to enhance pedestrians' safety by using sensors to detect their behaviors and alert vehicles to avoid potential risks (Malik et al., 2020). It complements vehicle-based active collision avoidance systems with direct communication between vehicles and pedestrians via various technologies (Kabil et al., 2022).

### 3.1.3. Data from the infrastructure

Besides the emerging vehicle-based sensors and communication technologies, data from the *infrastructure* are becoming increasingly important in supporting various functionalities of ITS (Li et al., 2020; Wang, 2018). Infrastructure-generated data support several key data-centric functions, including data collection and analysis, traffic management and control, safety enhancements, environmental monitoring, etc. (Zhang et al., 2011). *Data collection and analysis* relies on traffic detectors to continuously gather real-time traffic data, including vehicle flow, speed, density, and travel time. These data enable transportation authorities to have an accurate and up-to-date understanding of traffic conditions and form the foundation for data-driven decision-making and control (Sarker, 2022). *Traffic management and control* leverages rich data to adjust traffic controllers and adapt strategies to mitigate congestion and improve traffic flow (Guo et al., 2019). The data-centric approach ensures dynamic and responsive traffic management under changing conditions, including adaptive TSC, ramp metering control and variable message signs. Individual user-end devices (e.g., smart phones) could also be leveraged for traffic management by pushing dynamic information (e.g., road conditions, detours, and emergency alerts) to enhance road safety and driver awareness. By analyzing real-time traffic data and employing data analytics, the infrastructure can *enhance safety* by swiftly detecting traffic incidents such as accidents, road closures, or obstacles on the road. Incident data is communicated in real-time to emergency services, ensuring timely and efficient responses to accidents and emergencies (Lee et al., 2006). This rapid incident response, facilitated by data-centric processes, reduces congestion, and enhances road safety. Sensors on the infrastructure would be also responsible to *monitor environment*, such as air quality, temperature, humidity, and road surface conditions (Liu et al., 2023). This data-driven approach informs authorities about the environmental impact on road safety and allows for proactive responses. As electric vehicles (EVs) become increasingly prevalent, data from charging infrastructure also play an essential role in planning and operation of EV fleets (Wang et al., 2023c). As summarized by Shirvani et al. (2024), charging stations could be attacked by compromising their communication protocols connecting EVs and the power grid.

3.2. DPA against Vehicular Data

Section 3.1.1 provides an overview of the primary data sources that underpin the crucial functions of advanced vehicles, encompassing vehicle localization, perception, planning, and control. Notably, data originating from sensors, serving as the backbone of these functions, are of paramount importance, while a significant portion of existing DPAs against vehicles are directly related to these sensors (El-Rewini et al., 2020). Given the diverse operating principles of different sensors, an array of attack techniques is employed. A compact summary of attacks is given in Table 1, where existing DPAs targeting each key function are reviewed and categorized. In the following, we first introduce attacks following the categorization given in Table 1; more elaborations following the general framework will be provided in Section 4.

*3.2.1. Attacks targeting vehicle localization*

GNSS has long been recognized as susceptible to data attacks, with spoofing attacks standing out for their stealthiness and effectiveness. Unlike GNSS jamming, a typical denial-of-service attack, GNSS spoofing involves injecting false information into authentic GNSS measurements to deliberately mislead a vehicle's trajectory (Schmidt et al., 2016). These attacks can have significant consequences due to GNSS's crucial role in modern localization systems. Among GNSS spoofing attack strategies, the gradual drift attack has garnered significant attention for its stealthy nature (Wang et al., 2023a; Wang et al., 2021b; van Wyk et al., 2020). The gradual drift attack aims to modify a sequence of GNSS measurements over time to gradually divert the vehicle from its actual trajectory, resulting in a substantial deviation between the actual and falsified trajectories. Recent studies have proposed sophisticated, stealthy GNSS spoofing attacks that are challenging to detect. For example, a stealthy GNSS spoofing technique was designed to gradually shift the actual vehicle position according to its kinematic model (Wang et al., 2021b).

Besides attacks against GNSS, DPAs have also been developed for failing vision-based localization techniques such as SLAM. Wang et al. (2021) developed an easy-to-implement attack that ejects invisible light to introduce SLAM errors to AV without human notice. A stealthy attack, FusionRipper, was recently developed to target widely used Multi-Sensor Fusion algorithms that utilize multiple sensors (e.g., cameras, LiDAR and GNSS) for localization (Shen et al., 2020). The attack could disrupt production-grade autonomous driving systems, like Baidu's Apollo system.

*3.2.2. Attacks targeting vehicle perception*

Extensive research has exposed the vulnerabilities of sensors supporting vehicle perception to DPAs. Attacks on cameras and vehicles' computer vision algorithms are prevalent (Lu et al., 2017; Petit et al., 2015). For instance, DiPalma et al. (2021) devised an adversarial patch attack targeting camera-based obstacle detection. In this attack, an appropriately sized and designed adversarial patch is affixed to the rear of a box truck. The experiment evaluated this attack on an Apollo CAV operating in a production-grade autonomous driving simulator, showing that the vehicle under attack failed in detecting the box truck with the adversarial patch and thus collided into it. Like cameras, LiDAR data can also be manipulated with relative ease. The primary methods for attacking LiDAR include spoofing and relay attacks. A spoofing attack entails injecting signals into the LiDAR receivers of the target vehicles, while a relay attack involves using a transmitter and receiver to inject and receive signals from the target vehicles, respectively. For instance, Shin et al. (2017) introduced a delay component to defer the LiDAR signals returned from a target vehicle, with the delayed signals transmitted to the target vehicle by a malicious transmitter. Cao et al. (2019) demonstrated two attack scenarios: one involving an attack device positioned at the roadside emitting malicious laser pulses at passing vehicles, and another involving an attack device carried by a vehicle emitting malicious laser pulses at nearby victim vehicles. Petit et al. (2015) utilized two transceivers to relay LiDAR signals from the target vehicle to another at a different location. Furthermore, Yang et al. (2021) proposed an adversarial attack against DL models responsible for object detection using raw 3-D points from a LiDAR in a CAV.

Hacking into the waveform parameters of a millimeter radar sensor can disrupt its operation. Yan et al. (2016) conducted security experiments on a Tesla car's radar and autopilot system, revealing vulnerabilities to electromagnetic jamming and spoofing. Sun et al. (2021) conducted a comprehensive security analysis of millimeter-wave-based sensing systems in CAVs, revealing practical attack layers of millimeter radar sensors. They constructed multiple real-world attack scenarios to spoof victim vehicles.

Table 1. Summary of DPA on vehicular data

| Attacks | | Goal | Knowledge | Capability | | Strategy |
|---|---|---|---|---|---|---|
| | | | | Attack stage | Data alteration | |
| (Section 3.2.1) Attacks against Vehicle Localization | GNNS spoofing (Schmidt et al., 2016; Wang et al., 2023a) | Individual | White-box | Real-time decision | Feature perturbation | Dynamic system-based |
| | False data injection attack against SLAM (Wang et al., 2021a) | Individual | White-box | Real-time decision | Feature perturbation | Dynamic system-based |
| | Attacks on Multi-Sensor Fusion-based localization (Shen et al., 2020) | Individual | White-box | \ | Feature perturbation | Dynamic system-based |
| (Section 3.2.2) Attacks against Vehicle Perception | Spoofing attack on camera (Lu et al., 2017; Petit et al., 2015) | Individual | Black-box | Training | Label flapping | Complex data-driven |
| | Spoofing attack on LiDAR (Petit et al., 2015; Shin et al., 2017; Yang et al., 2021) | Individual | Black-box | Real-time decision | Feature perturbation | Complex data-driven |
| | Spoofing millimeter radar (Sun et al., 2021b) | Individual | White-box | \ | Feature perturbation | Dynamic system-based |
| (Section 3.2.3) Attacks against Vehicle Planning & Control | Attacks on vehicle trajectory planning (Deng et al., 2021) | Individual or Local | Black- or Grey-box | Real-time decision | Feature perturbation | Dynamic system-based |
| | False data injection attack against navigation systems (Cui et al., 2019) | Individual or Local | White-box | \ | Feature perturbation | Bi-level optimization |
| | Compromising AV's driving behavior (Cao et al., 2022) | Individual | Black-box | Real-time decision | Feature perturbation | Dynamic system-based |
| | Manipulating vehicle control (Wang et al., 2021c) | Individual or Local | Black- or Grey-box | Real-time decision | Feature perturbation | Dynamic system-based |

\* Symbol "\" stands for "not applicable"; the same for Table 2 and 3.

*3.2.3. Attacks targeting vehicle planning and control*

Vehicle planning and control relies on rich sensor data and information from accurate perception, including the vehicle's environment, real-time traffic conditions, environmental variables, and the vehicle's state (Vallati et al., 2021). Therefore, attacks against perception can lead to significant performance degradation in vehicle planning and control. For instance, a DPA can profoundly impact the accuracy of vehicles' detection modules, resulting in detrimental consequences such as planning an incorrect trajectory to avoid pedestrians (Chow and Liu, 2021).

Aside from indirect attacks failing vehicle localization and perception, DPAs can be directly aimed at compromising the planning and control of CAVs by manipulating critical information sources integral to these functions. For example, malicious interference may lead to less fuel-efficient planned trajectories at intersections, undermining energy efficiency (Ju et al., 2022). Moreover, adversaries could disrupt the harmonious merging of vehicles into platoons, disturbing CAVs' coordinated maneuvers (Zhao et al., 2022b). Attacks targeting traffic estimation and map databases can introduce inefficiencies in rerouting vehicles, thus undermining global planning strategies. For instance, Deng et al. (2021) devised an attack involving the insertion of adversarial triggers, such as square shapes or Apple logos, into the corners of original input images. Experimental results demonstrated that these malicious triggers in road images can significantly deviate the vehicle from its pre-planned trajectory. Through specific spatial-temporal driving behaviors, rather than physical or wireless access, the attacker can trigger actions that disrupt the planning and actions of the targeted vehicle. Notably, local map databases within in-vehicle networks, where each CAV maintains its map data, can be compromised by injecting malicious messages, impacting the accuracy and reliability of navigational information (Cui et al., 2019).

Moreover, attacks have also been directly applied to DL-based vehicle control. Wang et al. (2021) explored the backdooring of deep reinforcement learning-based AV controllers. Their methodology involves trigger design based on established traffic physics principles. These triggers encompass malicious actions such as vehicle deceleration and acceleration, capable of inducing stop-and-go traffic waves (congestion attacks) or causing the CAV to accelerate into the vehicle in front (insurance attacks). These attacks can be maliciously activated to provoke crashes or congestion when the corresponding triggers are detected. Cao et al. (2022) introduced an attack model that targets CAV's predictive driving behaviors. The attacker implicitly poisons the model by crafting an adversarial scenario involving surrounding agents to manipulate the CAV's behavior.

*Summary*: DPAs pose significant threats to every major data-intensive technology supporting CAVs. There are plenty of attacks on vehicle localization, among which threats to GNSS have long-been recognized. The emergence of data fusion-based localization systems, incorporating multiple sensors such as GNSS, IMU, LiDAR, and cameras, have drawn attention to attacks targeting these systems. Attack strategies are evolving: since conventional intuition-based strategies (e.g., instant noises for GNSS signal) are not effective for those data-fusion-based systems, recent studies identified more intricated attack strategies via carefully exploring and exploiting the weakness of these systems. Attacks on CAV's perception systems are mostly adapted from the cybersecurity domain (e.g., attacks on cameras) and customized to ITS data sources and application scenarios. Recent studies also investigated attacks on data fusion-based perception systems, such as computer vision systems deployed by combining camera and LiDAR data. On the other hand, direct attacks against CAV trajectory planning and control are limited, since attacks are primarily deployed by impacting localization and perception that could further influence CAV's planning and control outcomes. Direct attacks against vehicles' planning and control in driver-less scenarios likely pose higher threats than the indirect ones and thus warrant further investigation. Notably, most attacks on individual users are primarily focused on vehicles and are predominantly tested in simulations, with limited real-world implementation and validation.

3.3. DPA against V2X Communications

In scenarios involving V2X-supported connectivity, the impact of DPAs extends beyond individual vehicles and can affect other vehicles and road users. This section reviews attacks targeting V2V, V2I, and V2N. As noted earlier, attacks against pedestrians, including V2P, are lacking.

*3.3.1. Attacks targeting V2V*

DPAs against V2V are mainly focused on VANET applications to compromising the data confidentiality and integrity via Sybil attack, false data attack, and timing attack (Gao et al., 2022).

A *Sybil attack* refers to the act of creating multiple fake identities or nodes to gain control or manipulate a network or system. In a VANET, vehicles joining the network act as wireless nodes, and data are backed up among multiple nodes to enhance network availability. An attacker may use a single malicious node to impersonate multiple identities or propagate data backed up in the same malicious node. Similarly, malicious messages with multiple identities can be propagated to other nodes by the same malicious node. For example, an attacker may propagate a fake traffic scene to several nodes. When another normal node receives the fake traffic scene from those nodes, the normal node may modify its driving route. This may lead to a traffic accident (Guette and Ducourthial, 2007).

*False data attacks* forge identities to gain unauthorized access to the network, enabling them to alter or discard data packets transmitted in VANETs. An example of this type of attack is a malicious node disguising itself as an emergency vehicle to force other vehicles to slow down or stop (Mejri et al., 2014). By injecting false security information into the network or tampering with the broadcast security messages, attackers force legitimate vehicles to make choices that are not good for themselves, which might cause traffic accidents or increase traffic congestion on a certain road (Sumra et al., 2013). An attacker could also manage to fake sensor readings on its vehicle to create fake traffic messages and broadcasts them to the neighboring nodes to cause traffic jams (Al-kahtani, 2012). Zhao et al., (2022) for example, developed attack models that could affect vehicles' ramp merging behavior by falsifying the victim vehicle's BSM. The attacks were demonstrated likely to induce safety concerns and increase energy use.

A *timing attack* is a specific attack to delay the transmission of messages with high real-time requirements. As most messages with high real-time requirements are critical to the operation of a vehicle and the whole VANET, a malicious node in the network, which introduces abnormal latency to specific messages, is of great harm (Sumra et al., 2011). In some cases, malicious vehicles replace real-time information with that from a certain period in the past, failing systems relying on updated, real-time data. For example, a malicious vehicle could replay (broadcast) a previously transmitted message (e.g., about a traffic accident in the past) and use it to deceive other vehicles into believing fake traffic jams with the expired message (Mikki et al., 2013; Rawat et al., 2012). Such attacks are also termed as replay attacks.

### 3.3.2. Attacks targeting V2I

Attacks on V2I communication have been relatively limited, possibly due to the enhanced security measures in infrastructure components (Marcillo et al., 2022). Recently, there has been an increased focus on the vulnerability of connected vehicles-based TSC (or CV-TSC) relying on V2I communication. CV-TSC systems receive data from connected vehicles (CVs) through V2I communication, enhancing traffic signal optimization. However, this introduces the risk of compromised CVs transmitting poisoned data to manipulate signal timing plans, leading to suboptimal intersection operation (Feng et al., 2018). Multiple studies have focused on such indirect yet practical attacks on actuated and adaptive TSC, aiming to maximize system delay within constraints. Chen et al. (2018) investigated the threat that CV data spoofing by a single attack vehicle with the goal of inducing traffic congestion. It was shown that the attack could effectively reverse the benefits of CV-TSC systems, resulting in traffic mobility deteriorating by 23.4%. Yen et al. (2018) examined the impacts of time spoofing attacks on various TSC scheduling schemes, where CVs could manipulate their arrival times at intersections. Results from detailed simulations revealed that while the delay-based scheme exhibited superior fairness, it was more vulnerable to time spoofing attacks compared to other schemes. Irfan et al. (2022) proposed a Sybil attack where fake vehicles generate counterfeit BSMs, aiming to disrupt traffic signal timing and phasing changes. A RL-based attack model was developed, training an RL agent to determine an optimal rate of Sybil vehicle injection to induce congestion. Qu et al. (2021) formulated a novel task wherein a group of vehicles collaboratively sent falsified information to deceive RL-based TCS, which was shown effective in increasing vehicles' travel time. Recently, Feng et al. (2022) summarized the existing studies and proposed a comprehensive analysis framework for the cybersecurity problem in CV-TSC. With potential threats towards the major system components and their impacts on safety and efficiency, a data spoofing attack is considered the most plausible and realistic attack approach.

Besides targeting traffic management and control, attacks against V2I could also aim to CVs by manipulating RSUs that support vehicles via V2I. Rogue access points can be deployed along roads to mimic legitimate RSUs and launch attacks on associated users and vehicles (Han et al., 2014). Abhishek et al. (2021) introduced an attack model where adversaries compromise RSUs to disrupt the normal operation of CVs. After compromising an RSU, adversaries tamper with unencrypted data on the uplink and downlink of vehicles, affecting V2I communication. The primary goal is to cause economic losses or physical damage to vehicles and occupants.

*3.3.3. Attacks targeting V2N*

Attacks targeting *V2N* could compromise backend or commercial clouds that are often leveraged to handle large-scale transportation applications, particularly the operation of massive CVs. DPAs can compromise data confidentiality and integrity during V2N communication (Salek et al., 2022). Ristenpart et al. (2009) demonstrated that V2N are susceptible to attacks such as spoofing and Sybil attacks, which could risk CV data that are exchanged between vehicles and the cloud or stored in the cloud for further processing. In false information attacks, malicious actors manipulate information related to the surrounding environment transmitted from the cloud to CVs through V2N, introducing inaccuracies regarding the location and speed of nearby vehicles. Cybersecurity threats, such as information loss during transmission, identity spoofing, repudiation, information disclosure, and Sybil attacks, can compromise data integrity (Armbrust et al., 2010).

*Summary*: DPAs can maliciously manipulate the information exchanged via V2X, leading to incorrect decisions and potentially hazardous consequences. These attacks can target various aspects of V2X communication, such as identity spoofing, message integrity, and data injection. While attacks against V2V communication have garnered significant attention, particularly in relation to emerging V2V-supported applications like vehicle platooning, research on DPAs targeting V2I and V2N communication is relatively limited. Recent studies started to focus on V2I-supported CV-TSC, investigating their vulnerability to DPAs. The advent of V2N-supported ITS systems has gradually raised concerns about DPAs against V2N. By tampering with data exchanged between vehicles and data centers or roadside infrastructure, attackers can disrupt traffic flow across the network, compromise safety measures, and undermine the overall trustworthiness of the entire V2X system. On the other hand, DPAs against V2P are rarely discussed.

Table 2. Summary of DPA on V2X communications

| Attacks | | Goal | Knowledge | Capacity | | Strategy |
| --- | --- | --- | --- | --- | --- | --- |
| | | | | Attack stage | Data alteration | |
| (Section 3.3.1) V2V | Sybil attack to produce fake CVs (Guette and Ducourthial, 2007) | Individual or Local | White-box | \ | Feature Perturbation | Dynamic system-based |
| | False data attack against RSU-supported vehicles (Sumra et al., 2013; Zhao et al., 2022b) | Local | White-box | Training | Feature Perturbation | Dynamic system-based |
| | Timing attacks against VANET (Mikki et al., 2013; Rawat et al., 2012; Sumra et al., 2011) | Local | White-box | \ | Feature Perturbation | Dynamic system-based |
| (Section 3.3.2) V2I | False data attacks on CV-TSC (Feng et al., 2018; Huang et al., 2021a; Irfan et al., 2022; Yen et al., 2018) | Local | White- or Black-box | Training or Real-time decision | Feature Perturbation | Bi-level optimization |
| (Section 3.3.3) V2N | Spoofing and Sybil attacks against V2N-supported data exchanges (Ristenpart et al., 2009; Salek et al., 2022) | Network | White-box | \ | \ | Bi-level optimization |

Table 3. Summary of DPA on infrastructure data

| Attacks | | Goal | Knowledge | Capacity | | Strategy |
|---|---|---|---|---|---|---|
| | | | | Attack stage | Data alteration | |
| (Section 3.4.1) Traffic management & control | Attacks against fixed-time traffic control systems (Lopez et al., 2020; Perrine et al., 2019) | Local | White-box | \ | \ | Dynamic system-based |
| | Data falsify attack against ramp metering control (Ghafouri et al., 2016; Reilly et al., 2016) | Local or Network | White-box | \ | \ | Dynamic system-based |
| (Section 3.4.2) Data collection & analysis | DPA at collection (Prigg, 2014) or storage (Cao et al., 2019b; Vivek and Conner, 2022; Wang et al., 2018; Waniek et al., 2021; Zhao et al., 2021) | Network | White-box | Training | Feature perturbation | Bi-level optimization |
| | DPA against TSEP (Wang et al., 2024a), including traffic prediction (Liu et al., 2022; Zhu et al., 2023) | Local or Network | White- or Black-box | Training or Real-time decision | Feature perturbation | Bi-level optimization or Complex data-driven |
| (Section 3.4.3) Safety enhancements & charging infrastructure | False information attack against variable message signs (Kelarestaghi et al., 2018) | Local or Network | White-box | \ | \ | \ |
| | False information attack against RSU data (Hadded et al., 2020) | Local or Network | White-box | \ | \ | Dynamic system-based |
| | False data injection attacks on charging infrastructure (Acharya et al., 2020; Guo et al., 2023) | Network | White-box | \ | Feature perturbation | Dynamic system-based |

3.4. DPA against Infrastructure Data

DPAs against transportation infrastructure are limited to specific scenarios/applications. This section reviews attacks against infrastructure's several functions, including traffic management and control, data collection and analysis, safety enhancements, etc. (Table 3).

*3.4.1. Attacks targeting traffic management and control*

In TSC applications, attacks against the infrastructure directly (instead of indirect attacks via compromising CVs in Section 3.3), particularly controllers and detectors, can significantly impact traffic management and signal control. One approach involves modifying signal timing parameters. Ghena et al. (2014) examined an attack model in which an adversary directly alters the signal timing parameters of controllers connected to the central management system through a wireless network. The attacker gains partial control of TSC through means like debugging ports of the operating system or the National Transportation Communications for ITS Protocol commands (US). Another attack avenue involves the modification of green time ratios, as introduced in Lopez et al. (2020). This model seeks to alter the green time ratios of fixed-time traffic control systems. The impact of this approach was evaluated in the context of intersections and grid networks in Manhattan, showing that modifying the green time ratios could substantially reduce the flow rate. Similarly, Perrine et al. (2019) proposed district-wide attacks that prioritize intersections with the maximum traffic flow or those potentially affecting the most vehicles. The results revealed a significant increase in the total traffic delay in the district when multiple signals are targeted.

In an alternative strategy, Ghafouri et al. (2016) proposed to perturb traffic flow sensor data, which provides input to fixed-time controllers for signalized intersections. The attacker formulates a bilevel programming optimization problem to help identify sensor perturbations. The results revealed that compromising only a small number of sensors can lead to severe network congestion.

Lastly, ramp metering control attacks, studied by Reilly et al. (2016), target the metering control mechanisms on freeways. These attacks aim to generate complex congestion patterns, with objectives ranging from causing network-wide congestion to facilitating an escape during a police pursuit.

*3.4.2. Attacks targeting data collection and analysis*

Infrastructure in ITS often plays a pivotal role in collecting, analyzing, and distributing traffic data, making it a target for DPAs. These attacks can manipulate data at various stages, such as the data collection, storage, and analysis stages. At the data *collection* stage, one attack avenue targets wireless detectors, which use technologies like Bluetooth and Wi-Fi to detect travelers. Compared to traditional loop detectors, these detectors are considered more vulnerable. Attackers can fake arbitrary traffic data, compromising the information collected from these detectors. This manipulation can disrupt traffic monitoring and impact control decisions (Prigg, 2014).

Data at the *storage* stage is also vulnerable to DPAs, particularly concerning crowdsourced mapping and navigation that rely on extensive location data (Sinai et al., 2014). Sybil attacks have been highlighted as an effective method where attackers compromise the data aggregation system by creating and injecting numerous fake identities. The poisoned datasets can report false traffic conditions, leading to rerouting decisions by vehicles to circumvent these false jams, ultimately reducing efficiency (Cao et al., 2019b; Wang et al., 2018; Waniek et al., 2021). Zhao et al. (2021) developed a DPA model to fake users' location records, thereby enabling manipulation of the aggregated mobility network. This attack is formulated as a min-max optimization problem and has been demonstrated to pose a significant threat to data aggregation and mobility network derivation.

Traffic State Estimation and Prediction (TSEP) serves as one of the major purposes of data *analysis*. Spatiotemporal based TSEP models, especially those based on ML and DL are gaining popularity. Such models serve as data infrastructure to support key ITS functions, such as travelers' daily navigations. Attacks have been recently proposed and evaluated to compromise these models. Wang et al. (2024) developed a DPA model targeting TSEP. The attack model was demonstrated effective in deviating queue length estimation and compromising vehicle classifications. Liu et al. (2022) examined the vulnerability of these models and introduced an adversarial attack framework that targets specific data sensors to degrade the performance of traffic predictions. Similarly, Zhu et al. (2023) introduced the concept of diffusion attacks to assess the robustness of traffic prediction models based on Graph Convolutional Networks. The attack selects and poisons nodes in the graph network, and testing across two cities showed the effectiveness of these attacks.

*3.4.3. Attacks targeting infrastructure for safety and other functions*

Infrastructure data for critical safety applications could also be vulnerable to DPAs. It has been reported that attackers can modify variable message signs to corrupt the safety and efficiency trips (Kelarestaghi et al., 2018). Comert et al. (2021) demonstrated false information attacks that modify BSM from RSUs for manipulating vehicle behaviors and generating incorrect positions. The vulnerability of RSU supporting collective perception-based on-ramp merging control has also been reported (Hadded et al., 2020).

With the growing adoption of EVs, ensuring the security of charging infrastructure is of paramount. Basnet and Ali (2021) highlighted the presence of potential backdoors in charging infrastructure, underscoring the importance of addressing these vulnerabilities to ensure system security. Acharya et al. (2020) investigated the impact of false data injection attacks on charging infrastructure, revealing that such attacks can lead to significant delays and system oscillations, potentially disrupting charging services. Guo et al. (2023) explored delayed charge attacks, an insidious method capable of disrupting EV charging services, resulting in increased queuing times, unfulfilled requests, and revenue loss. Collectively, these studies emphasize the critical need to enhance the security of EV charging infrastructure.

*Summary*: While infrastructure data plays a crucial role in various applications, the investigation of threats posed by data attacks remains limited. The threats to infrastructure data are mostly focused on the data collection systems (e.g., detectors), data storage (e.g., map databases), variable message signs and TSC. Recently, DPAs have been developed for new types of infrastructure, such as wireless detectors, RSUs, and charging infrastructure. Additionally, emerging TSEP applications that heavily rely on infrastructure data have been found vulnerable to DPAs. As intelligent infrastructure emerges as a key pillar supporting ITS applications, more attention should be devoted to investigate its susceptibility to DPAs.

## 4. General Framework

Here we discuss and synthesize the reviewed DPAs in ITS by following the general framework in Figure 2. The focus is on the four major components of an attack model, as well as on the risk assessment of an attack. The summaries are also included in Table 1-Table 3.

4.1. Attack Goal

As shown in Table 1-3, DPAs in ITS exhibit diversity in terms of their goals. Considering the extensive range of application scenarios, the attack goals can be categorized from various perspectives, including data types and sources, the type of security breach caused by the attack (e.g., integrity violation), and transportation metrics (e.g., safety and efficiency). Here, we present a summary of attack goals in ITS, classified into three groups based on the scale of their potential influences: individual road users, local corridors, and network-wide consequences.

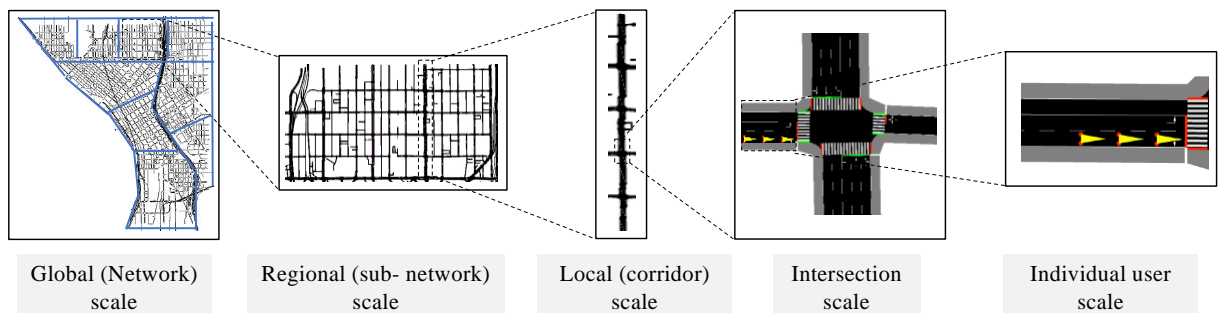

Figure 4. Illustration of the ITS operation at multiple scales. (From Guo et al. (2019))

As highlighted in Guo et al. (2023), ITS operations span multiple scales within their spatial and temporal domains, encompassing the movement of individual vehicles, the flow within intersections or freeway segments, the dynamics of corridors and subnetworks, and the behavior of an overall network (Figure 4). Each operational scale is characterized by distinct behavioral dynamics while interconnected with operations at other scales. Therefore, an attacker can influence one particular scale by targeting this scale directly or another interacting scale indirectly. Here we introduce the three scales of attack goals and provide the categorization of attacks in Table 1-Table 3.

*Influence on individual users:* Influence on individual users represents one common attack goal within ITS. Attacks could target a specific vehicle and aim to disrupt the normal operation of the victim. As shown in Table 1, the attacks can achieve the individual-level goal by compromising systems deployed on individual vehicles, such as GNNS spoofing targeting localization (Wang et al., 2023a), camera spoofing targeting perception, or vehicle trajectory manipulation targeting vehicle planning and control. This would affect the safety, fuel efficiency, or overall performance of the victim vehicle. Alternatively, attackers might indirectly influence the individual victim, for example, with Sybil attack to produce fake CVs and interact with the victim (Guette and Ducourthial, 2007), or disrupting the surrounding environment estimation by deceiving neighboring vehicles (Table 2).

*Local influence:* Instead of a specific victim, attackers could focus on the efficiency and safety of a local corridor or area. For instance, attacks against V2V could create indiscriminate disruptions, potentially influencing multiple vehicles on a corridor (Guette and Ducourthial, 2007). Attacks targeting V2X and infrastructure data are likely to create local influence, as shown in Table 2 and Table 3. Examples include attacks against CV-TSC, where manipulating input data, often sourced from CVs, can compromise intersection efficiency without specific personal benefits (Feng et al., 2018; Huang et al., 2021a; Irfan et al., 2022; Yen et al., 2018). Emerging technologies like CAVs acting as traffic controllers for enhancing corridor stability and safety are also subject to manipulation (Guette and Ducourthial, 2007). Attackers may aim to destabilize vehicle platooning via timing attacks, leading to safety issues on the corridor (Mikki et al., 2013; Rawat et al., 2012; Sumra et al., 2011). Road-side infrastructure on a corridor, including sensors supporting vehicle localization, perception, and environmental monitoring, can be attacked to create local influence (Sumra et al., 2013; Zhao et al., 2022b). Table 3 shows that ramp metering controllers for local corridors are also common targets (Ghafouri et al., 2016; Reilly et al., 2016). Attacks on these controllers result in traffic congestion in the corridor, affecting multiple road users.

*Network-wide influence:* The attacker would also try to compromise transportation performance at a network level. As shown in Table 3, these attacks often target infrastructure data that support TSEP, data aggregation systems, charging stations, and cloud-supported CVs. For instance, attackers may poison data collection (e.g., for digital map databases) to manipulate traffic state estimation, leading to the rerouting of vehicles, suboptimal trajectories, and lower network-wide efficiency (Prigg, 2014). Given their importance to navigating massive vehicle, network-wide forecasting models trained on storage data can also be compromised to generate network-wide influence (Cao et al., 2019b; Vivek and Conner, 2022; Wang et al., 2018; Waniek et al., 2021; Zhao et al., 2021). Similarly, attackers might also disseminate misinformation on social networks to misguide travelers in the road network. Network-level data aggregators may become the targets, as shown by Zhao et al. (2021), where human mobility measures used for network planning can be manipulated by falsifying human trajectories. Recent research has also identified the charging infrastructure as a viable target to disrupt system-level operations, such as reducing the operational efficiency of EV-based taxi fleets (Acharya et al., 2020; Guo et al., 2023). Besides targeting the infrastructure, Table 2 shows that spoofing or Sybil attacks against V2N that support CAVs could also induce network-level influences (Ristenpart et al., 2009; Salek et al., 2022).

4.2. Attacker's Knowledge

As shown in Table 1-Table 3, most DPAs in ITS assume a white-box attack scenario, i.e., attackers have full knowledge of the underlying model and data, necessitating substantial knowledge about the victim systems. This knowledge can span a broad spectrum due to the complexity of ITS, encompassing hardware, software, communication channels, and the interactions among subsystems. Such setting means attackers must possess information about the data generation mechanisms of sensors (Wang et al., 2021a; Table 1), learning algorithms for CV-TSC (Huang et al., 2021a), V2X communication protocols (Guette and Ducourthial, 2007; Table 2), and domain knowledge concerning various aspects of transportation, such as traffic rules-aware traffic control (Ghafouri et al., 2016; Reilly et al., 2016; Table 3) and traffic dynamic-based prediction (Liu et al., 2022; Zhu et al., 2023; Table 3). Attacks targeting TSC, for instance, often require comprehensive knowledge of the TSC system, including control algorithms, model parameters, and implementation specifics (Lopez et al., 2020; Perrine et al., 2019). Some studies even assume that attackers can directly manipulate open and accessible attack surfaces, a subset of white-box attacks. While white-box attacks can demand extremely high knowledge and make certain assumptions, they remain valuable as they help identify critical vulnerabilities and inform countermeasure prioritization.

In contrast, black-box attacks are relatively rare currently in DPAs in ITS (see Table 1-3) but are emerging as an active research direction, as they require limited to no prior knowledge of the victim. For example, a recent study introduced a black-box attack scenario targeting CV-TSC systems, with the initial step involving learning the control logic through queries and subsequently launching attacks based on this learned logic (Huang et al., 2021a). Black-box attacks have become a common assumption in the cybersecurity domain due to the wide deployment of DL methods for which it is challenging for the attackers to have the full knowledge of the model structure or parameters. We expect that this concept may also be gradually applied to the study of DPAs in ITS. For example, attacks on vehicular computer vision systems (Table 1), including cameras (Lu et al., 2017) and LiDAR (Yang et al., 2021), often adopt black-box attack models. Attacks that apply DL models such as CAV trajectory planning and control (Deng et al., 2021; Table 1) and traffic prediction (Liu et al., 2022; Zhu et al., 2023; Table 3) are also designed in black-box scenarios. There are also grey-box attacks that are in between white-box and black-box attacks, where partial knowledge of target model and data is required. Yet, most existing attack models are either in white-box or black-box scenarios and rarely in gray-box attack scenarios (Deng et al., 2021; Wang et al., 2021c). Given the diversity of real-world ITS applications, it is expected that gray-box attacks may emerge in future studies to reflect certain practical settings.

4.3. Attacker's Capability

Depending on an attacker's access to data, the attacker may launch attacks by altering a subset of the dataset at the training or real-time decision stage. The attacker's capability is also reflected by the way the data is altered. As shown in Table 1-Table 3, some attacks in ITS do not discuss attackers' capability, assuming that all required resources are available to attackers to launch the desired attack.

*Influence on the stage of data utilization:* If a system relies on a model or algorithm trained/learned on data, DPAs can occur during two distinct phases: the real-time decision (i.e., online inference) or the training stage. Most attacks in ITS take place during the inference stage when the trained model is used to support *real-time decisions*. Attacks may involve interfering with data sources directly, such as manipulating GNSS signals (Schmidt et al., 2016; Wang et al., 2023a) or other sensor data (Table 1), which can compromise the normal operations of victim systems. Alternatively, attackers with access to data collection during the inference stage may poison the data at its source (e.g., attacks against traffic prediction (Liu et al., 2022; Zhu et al., 2023)). As more ITS systems adopt ML or DL models that are pre-trained using extensive datasets, attacks may also target the *training stage*. Such attacks often involve manipulating a subset of training data, as demonstrated in previous research (e.g., attacks against vehicular computer vision systems (Lu et al., 2017; Petit et al., 2015) and vehicle classification system (Wang et al., 2024a)). This results in a compromised model, negatively impacting the subsequent inference stage.

*Data alteration:* Constrained by a certain budget, the portion of data being manipulated could be limited. It is often assumed that an attacker alters either data features or data labels (e.g., label-flip). Altering feature values and flipping labels of samples represent distinct methods of DPAs with unique implications. When feature values are altered in ITS datasets, it means manipulating information related to road conditions, traffic behavior, or sensor readings. This can lead to biased traffic predictions (Liu et al., 2022; Zhu et al., 2023) or unsafe route navigation (Sinai et al., 2014; Zhao et al., 2021), directly impacting the effectiveness of ITS systems such as traffic management and route planning. On the other hand, flipping labels, though relatively rare in ITS data (as shown in Table 1-Table 3), involves changing the ground truth information about traffic situations or vehicle trajectories. This can mislead the learning process of ITS models, resulting in incorrect object detection and safety concerns for CAVs (Lu et al., 2017; Petit et al., 2015; Table 1).

4.4. Attack Strategy

In general, a DPA aims to find an optimal set of perturbations to the pristine data to achieve certain attack goal when the data are used by ITS systems (certain learning models in particular). The attack strategy defines how the attacker manipulates data to launch the DPA, which are directed by the attack goal and constrained by the attacker's knowledge and capabilities, as well as application-specific limitations. Consequently, identifying an optimal attack strategy can be challenging.

As shown in Table 1-Table 3, DPA strategies in ITS can be diverse, due to the broad spectrum of ITS applications. We can unify and formulate the strategies as a bi-level model below:

$$\max_{\delta} O(V, M, \theta^*) \quad \text{(1-a)}$$

s.t. $\delta \in \Delta$, and other possible constraints on the perturbed data  (1)

$$\theta^* \in S(D_c \cup D_p^\delta, M, \theta) \quad \text{(1-b)}$$

Here, the lower-level problem (1-b) represents a specific ITS application $S$, where the solution $\theta^*$ is derived from a set of input data using a versatile learning model $M$. $M$ can encompass various ML and DL models. $D_c$ and $D_p^\delta$ stand for the pristine input data and the poisoned data, respectively, following the data manipulation $\delta$. $D_p^\delta$ could be a portion of the dataset that is perturbed or a set of poison data that is added to the pristine dataset. $\delta$ is restricted within a specific feasible region $\Delta$ and may include other problem-specific constraints. $\Delta$ and the problem-specific constraints represent the attacker's knowledge and capability, which should be formulated based on the specific ITS models and attack scenarios. Conventionally, the perturbation magnitude of $\delta$ is confined with $\|\delta\| \leq \epsilon$ (a small positive value) to ensure the DPA is *stealthy*. The upper-level problem (1-a) is to maximize the attacker's goal, represented by an objective function $O$, which relies on the solution $\theta^*$ of the ITS model, learning model $M$, and validation data $V$ (or other data for evaluating attack performance). Below, we show how this general attack strategy can be tailored for specific ITS learning models, including optimization-based ITS models, ITS models represented by general dynamic systems, and other intricate data-driven ITS models (see summary in Figure 5). Table 1-Table 3 categorize each reviewed attacks according to their strategies.

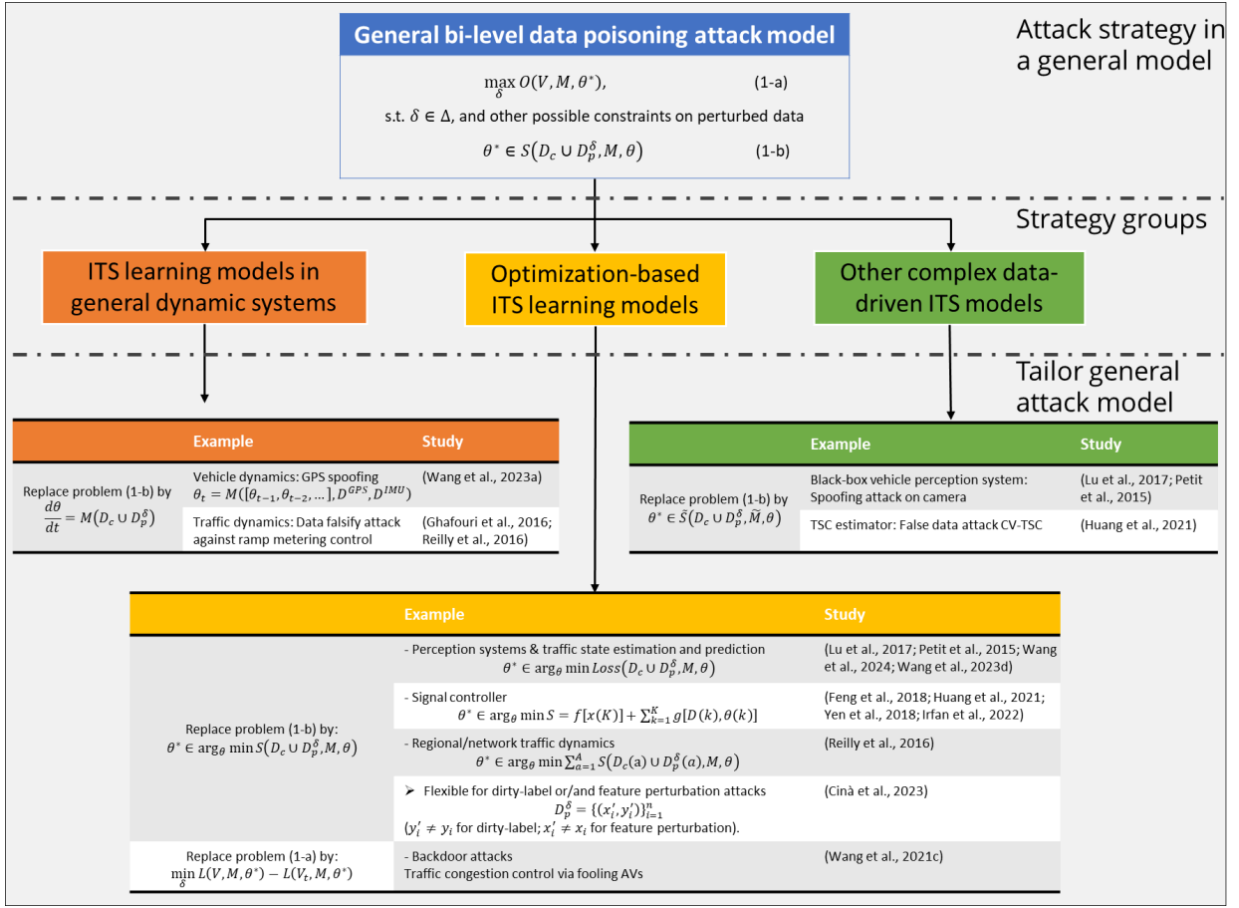

Figure 5. The general and specific formulations of attack strategy

*4.4.1. Optimization-based ITS learning models*

For vehicle perception and TSEP (Wang et al., 2024a; Wang et al., 2023d), the lower-level problem (1-b) can often *be framed as an optimization problem*, where a solution is learned from a dataset:

$$\theta^* \in \arg_\theta \min S(D_c \cup D_p^\delta, M, \theta). \qquad (2)$$

Many ITS attacks can be formulated by substituting problem (1-b) with (2), with an additional step of defining the attacker's objective function. As a result, the attack strategy is represented by a *bi-level optimization problem*. As summarized in Table 1-Table 3, this strategy has been applied to spoofing attack on camera (Lu et al., 2017; Petit et al., 2015), false data injection attack against navigation systems (Cui et al., 2019), false data attack against CV-TSC (Feng et al., 2018; Huang et al., 2021a; Irfan et al., 2022; Yen et al., 2018), poisoning attacks against V2N supported CVs (Ristenpart et al., 2009; Salek et al., 2022), spoofing attack against TSEP (Wang et al., 2024a), attacks against data at collection or storage (Cao et al., 2019b; Sinai et al., 2014; Wang et al., 2018; Waniek et al., 2021; Zhao et al., 2021), poison traffic predictor (Liu et al., 2022; Zhu et al., 2023), etc.

For instance, consider an attack against a vehicle classification system (Wang et al., 2024a) (see Table 3). Model $M$ could be an ML or DL model (e.g., SVM) trained offline on a set of vehicle features (e.g., velocity and acceleration), which can be maliciously manipulated through $D_p^\delta$:

$$\theta^* \in \arg_\theta \min Loss\big(D_c \cup D_p^\delta, M, \theta\big). \tag{3}$$

Here, $\theta^*$ represents the learned model parameters evaluated on a validation dataset. In this context, the objective function in the upper-level problem $O(V, M, \theta^*)$ is the loss function (e.g., L2 norm) computed on the validation dataset $V$. The attacker seeks to optimize the perturbation $\delta$ (applied to the poisoning samples $D_p^\delta$) to maximize the loss function, i.e., to increase the classification error of the target model $M$ on the clean validation samples $V$. Similar strategy can be applied by adapting the general attack model to other systems that rely on specific data learning models, including a camera-based vision system for traffic sign classification on CAVs trained on a set of images. Solving the bi-level problem to identify the optimal perturbation can be challenging. Existing solutions often rely on gradient-based methods to find an optimal set of perturbations by evaluating the gradient of the adversarial objective over data perturbations (i.e., how the former changes with the latter). However, as shown in Wang et al. (2024) and detailed in Section 5.1 (challenge of solving the attack model), the gradient-based solutions have limitations in ITS applications. Some promising directions (e.g., semi-derivative-based method) to address the limitations will be discussed in Section 6.1.

Signal control is another notable application of bi-level optimization-based model. Guo et al. (2019) formulated a general arterial traffic control problem as an optimization problem, where the traffic state and environment variables serve as input data $V$. The decision variables are equivalent to the solution $\theta^*$:

$$\theta^* \in \arg_\theta \min S = f[x(K)] + \sum_{k=1}^{K} g\big[D_c(k) \cup D_p^\delta(k), \theta(k)\big] \tag{4}$$

*s.t. Initial conditions, traffic flow dynamics, & vehicle dynamics.*

The objective of signal control is to optimize a specific performance index $S$ over a finite time horizon $[0, K]$. $S$ typically comprises mobility-based or sustainability-based objectives or a combination of the two, while $f[\cdot]$ and $g[\cdot]$ represent certain functions. The decision variables are a sequence of control inputs $\theta^* = [\theta^*(1), \ldots, \theta^*(K)]$. This general formulation can be adapted to various DPAs against CV-TSC by replacing problem (2) with problem (4), and specifying the attacker's objective function (e.g., maximizing travel delays at the targeted intersections).

In cases that the optimization problem (4) is represented by a network-level traffic state estimation model, the bi-level optimization-based attack strategy could create network-level influence. For instance, it could be an ML or DL model trained on traffic data for future traffic prediction (Liu et al., 2022; Zhu et al., 2023; see Table 3). In this case, $M$ and $\theta^*$ collectively represent a learned, optimal traffic predictor derived by minimizing training errors. By specifying the attack objective in problem (1-a) as the prediction errors (in the reference stage), the attacker aims to identify a set of perturbations $D_p^\delta$ that could poison the traffic predictor and increase network-wide prediction errors. Alternatively, the ITS learning model could be:

$$\theta^* \in \arg_\theta \min \sum_{a=1}^{A} S\big(D_c(a) \cup D_p^\delta(a), M, \theta\big). \tag{5}$$

Here, the summation of all sub-areas $a = [1, \ldots, A]$ evaluates the total influence on the network.

The general model can also accommodate attackers' capability to tamper with the dataset. For instance, the attack model is flexible to represent *label-flipping* or *feature perturbation* attacks or both. Define $D_p^\delta = \{(x_i', y_i')\}_{i=1}^n$, with $x_i'$ being the poisoning feature values and $y_i'$ being the poisoning label implemented by the attacker. $(x_i, y_i)$ is the pristine data and $n$ is the number of poisoned data samples. Then, $y_i' \neq y_i$ in the lower-level problem (1-b) represents a label-flipping attack and $x_i' \neq x_i$ represents a feature perturbation attack. If $y_i' \neq y_i$ and $x_i' \neq x_i$, both the feature values and data labels are perturbed.

Though rare in ITS (Wang et al., 2021c), backdoor attacks can also be handled by the general framework (Figure 5). During model training, a backdoor attack implants a backdoor by introducing some patterns (known as trigger) into the poisoning samples. During the inference stage, the samples containing the same trigger can activate the backdoor attack. In order to design feasible and stealthy backdoor triggers, we can formulate the backdoor poisoning attack as a bi-level optimization problem that learns the trigger by modifying the adversarial objective in Problem (1) (Cinà et al., 2023; Wenger et al., 2021).

$$\min_\delta L(V, M, \theta^*) - L(V_t, M, \theta^*)$$
$$\text{s.t. } \delta \in \Delta, \text{ and other possible constraints on perturbed data} \tag{6}$$
$$\theta^* \in L\big(D_c \cup D_p^\delta, M, \theta\big)$$

Here, the attacker optimizes the perturbation $\delta$ for poisoned samples in $D_p^\delta$ to mislead the performance of $M$ on a set of samples with trigger $t$ ($V_t$), while preserving the accuracy on $V$ to make the attack stealthy. More details about backdoor attacks, especially their potential in ITS applications, are introduced in Section 6.4.

*4.4.2. ITS models represented by general dynamic systems*

Some ITS systems do not fit into an optimization framework but can be represented as a dynamic system $M$, as expressed in Equation (7) below. Such attack strategy involving dynamic systems has been widely applied for attacking all the primary ITS data sources (Table 1-Table 3).

$$\frac{d\theta}{dt} = M(D_c \cup D_p^\delta) \tag{7}$$

In this scenario, the lower-level problem (1-b) could comprise a set of time-variant equations describing how system variables change over time. These variables may encompass vehicle speed, traffic flow, congestion levels, signal timings, vehicle trajectories, and environmental factors like weather conditions. Such equations should capture the cause-and-effect relationships between components, including vehicles, infrastructure (e.g., traffic signals, RSUs), and communication networks. The solution $\theta^*$ may denote vehicle or traffic status derived from the system dynamics and input data.

As shown in Figure 5, while attacking systems that interact with traffic, traffic dynamics can be modeled as dynamical systems. For instance, when targeting metering controllers, it is necessary to model the dynamics of traffic flows and how they interact with the controllers. An attack can then influence traffic patterns (e.g., creating congestions) by manipulating the controllers (Ghafouri et al., 2016; Reilly et al., 2016). Another example is GNNS spoofing attacks, in which Problem (7) can be an Extended Kalman Filter-based localization model that tracks vehicle dynamics by fusing data from multiple sensors (e.g., IMU and GNNS in Wang et al., 2023a; Table 1). Omitting the details, the localization model can be expressed below.

$$\theta_t = M([\theta_{t-1}, \theta_{t-2}, \ldots], D^{GPS}, D^{IMU}) \tag{8}$$

Here, $\theta_t$ and $[\theta_{t-1}, \theta_{t-2}, \ldots]$ stand for the vehicle's current and historical positions. Function $M(\bullet)$ denotes localization model. $D^{GPS}$ and $D^{IMU}$ provide data from GNNS and IMU, respectively. The similar strategy applies to vehicle dynamics such as CVs' adaptive cruise control (Guette and Ducourthial, 2007; see Table 2), and routing decisions (Sinai et al., 2014).

*4.4.3. Other complex data-driven ITS models*

In scenarios where a data-driven ITS system is complex for attackers, it can be challenging to express it as an optimization model or explicitly set up a dynamic system model. As shown in Table 1-Table 3, attack strategies suitable for such scenarios are rare in existing studies. However, such scenarios would emerge with the fast evolution of ITS, which likely leads to advanced systems. For example, modern vehicle localization systems could involve multiple techniques including the LiDAR-based localizations, making it impossible to explicitly model their response to data changes (the core for bi-level attack model) (Yang et al., 2021; see Table 1). In such cases, we can treat the lower problem (1-b) as a black-box and attempt to approximate it with an estimator function (Ilyas et al., 2018; Liu et al., 2022).

$$\theta^* \in \tilde{S}(D_c \cup D_p^\delta, M, \theta) \tag{9}$$

The estimator function $\tilde{S}$ approximates the complex objective function, allowing the attacker to use it to identify the attack strategy. The approximation method can rely on learning models to learn the estimator functions from data. For example, as summarized in Table 2, Huang et al. (2021) assumed that the control logic of a CV-TSC system is too complex for an attacker to manipulate. The attacker then learns the control logic using a surrogate model and launches falsified data attacks to influence TSC performance.

In cases of *black-box attacks*, the data and learning models are replaced by surrogate data $D'_c$ and a surrogate model $M'$. Here, the attacker is assumed to collect a surrogate dataset (e.g., via queries) and use it to train a surrogate model that approximates the target. Strategies applicable to white-box attacks can then be derived using the surrogate model and subsequently transferred against the original target model $M$. This idea relies on the so called *transferability* of DL attacks (Xie et al., 2022), which is under active explorations in ITS (Wang et al., 2024b). In some cases, attackers may have knowledge of an ITS system but prefer to obtain estimator functions with better properties (e.g., smoothness). For instance, optimizing a non-differentiable loss can be challenging, and thus using a smoothed version might be more effective (Cinà et al., 2023). The resultant attack strategy is then applied to mislead the unknown or complex victim system.

4.5. Risk Assessment

Definitions or measures of impact and likelihood related to transportation cybersecurity are lacking. Feng et al. (2022) identified the impact measures such as individual travel time/delay (individual goal), the number of potential crashes/conflicts (safety goal), total delays in the network (mobility goal), or a combination of those measures. To facilitate our discussion, this study follows the definitions of impact and likelihood of certain risk in the National Institute of Standards and Technology (Stoneburner et al., 2002), which defines the key measures and three (i.e., high, medium and low) levels (Table 4).

As mentioned in Section 2.2, due to the challenge in real-world evaluation and data collection, studies systematically evaluating the impact and likelihood of DPAs in ITS are currently sparse but are gaining increasing attention (Feng et al., 2022). This section provides discussions on how the proposed attack framework generates insights to address the challenge. They are qualitative and conceptual and are based on the insights obtained from the proposed attack framework, which relies on the evolving, common understanding of DPAs. Rigorous validations and real-world data support on quantitative assessments of the impacts and likelihood are critically needed. Concise summaries are presented in Tables 5-7 in Appendix A.2, with tentative results on impact and likelihood assessment. The results in Tables 5-7 are subject to change when quantitative evaluations are done with real-world data. More discussions on the challenges in risk assessment of DPAs in ITS can be found in Section 5.3, followed by future research needs (e.g., quantitative and more objective assessment) in Section 6.3.

Table 4. Definitions of attack impact and likelihood (Stoneburner et al., 2002)

| Impact Magnitude Definitions | |
|---|---|
| **Magnitude** | **Definitions** |
| High | Attacks may result in the highly costly loss of major tangible assets, lead to human death or serious injury, or significantly violate/harm an organization's mission, reputation, or interests. |
| Medium | Attacks may result in costly loss of tangible assets, or result in human injury or violate/ harm an organization's mission, reputation, or interests. |
| Low | Attacks may result in the loss of some tangible assets or noticeably affect an organization's mission, reputation, or interests. |
| Likelihood Definitions | |
| **Levels** | **Definitions** |
| High | The attack is highly motivated and capable, with ineffective controls to prevent it. |
| Medium | The attack is motivated and capable, but controls are in place that may impede its successful execution. |
| Low | The attack lacks motivation or capability, or effective controls are in place to prevent or significantly impede its execution. |

*4.5.1. Evaluating attack impacts*

It is important to note that the impact of an attack can vary even when they share the same level of influence in their attack goals. For example, attacks aimed at influencing *individual users* may have a broad range of impacts, from mild disturbances in driving reactions to severe risks to users' lives (Table 5). Specifically, GNSS spoofing has demonstrated the potential to cause accidents by diverting vehicles from safe paths (Schmidt et al., 2016; Wang et al., 2023a). Given their critical roles in supporting advanced vehicles and preventing accidents, successful attacks on cameras and radars could lead to high impacts (Lu et al., 2017; Petit et al., 2015). Misclassification or misinterpretation of road objectives due to poisoning of LiDAR data would disrupt drivers' or driving systems' actions (Petit et al., 2015; Shin et al., 2017; Yang et al., 2021). On the other hand, in real-world implementations, LiDAR is often coupled with camera-based computer-vision systems to cross-validate data, and thus the complimentary role makes its failure a medium impact. Ultrasonic sensors, often used in low-speed scenarios like parking assist, may have a low impact when attacked (Sun et al., 2021b). Attacks against vehicle planning and control typically result in high impacts due to their direct interference on vehicle's driving systems, endangering road users' safety.

When attacks aim to create *local influence*, their impacts may also vary (Table 6). Attacks that compromise V2V may disrupt cooperative vehicles but with medium impacts (Sumra et al., 2013; Zhao et al., 2022b): aside data from V2V, individual vehicles often rely on redundant systems to support safety-critical functions, mitigating the impact. For instance, attacks against platooning may reduce the stability and efficiency of the platoon but are difficult to result in crashes due to the presence of emergency braking systems on individual vehicles (Mikki et al., 2013; Rawat et al., 2012; Sumra et al., 2011). On the other hand, false data attacks against CV-TSC could have high impacts, as prior studies have demonstrated that the attacks could significantly reduce efficiency and compromise safety (Feng et al., 2018; Huang et al., 2021a; Irfan et al., 2022; Yen et al., 2018). Attacks on V2N also likely have high impacts, given the large number of victims and network-wide disruptions (Ristenpart et al., 2009; Salek et al., 2022).

*Network-level* attacks can be extremely costly to conduct but typically have high impacts due to their large scale (Table 7). For example, fake traffic jams created by attackers can induce disproportionately large disruptions in the road network, due to travelers' adaptive behaviors (Cao et al., 2019b). Vivek and Conner (2022) demonstrated that by identifying the most critical road segments in the Boston network, attackers could trigger cascading traffic jams that significantly reduce network efficiency. Disinformation attacks that divert traffic to a central area can cause network-wide disruptions and threaten safety operations, such as emergency responses, in congested areas (Waniek et al., 2021). High impacts can also result from attacks on network-level spatiotemporal traffic prediction models, where many travelers may experience additional delays due to misleading navigation based on predictive travel time (Liu et al., 2022; Zhu et al., 2023). Guo et al. (2023) proposed and tested a delay charging attack model using real-world taxi trip data and EV supply equipment locations in New York City. The results show that a 10-minute delay attack could lead to 12-minute longer queuing times and 8% more unfulfilled requests, resulting in an 11% weekly revenue loss per driver. Some infrastructure may have embedded safety mechanisms in response to unexpected failures, which could mitigate attack impacts. For instance, attacks against fixed-time traffic control systems have medium impacts, as the attack controllers would switch it to flashing red lights if there is malfunction that induce conflicts impacting road users' safety (Lopez et al., 2020; Perrine et al., 2019).

*4.5.2. Evaluating likelihood*

The general attack framework can also aid in assessing the likelihood of successfully executing these attacks, including the frequency of specific attack and the probability of success once the attack is launched. As illustrated in Figure 2, the categories that account for attacker's knowledge, capability, and attack strategy provide a basis for evaluating the attack likelihood, which is elaborated below.

*Attacker's knowledge*: White-box attacks, where adversaries possess comprehensive knowledge of the system's architecture and algorithms and can iteratively refine their attacks based on feedback, are often more likely to succeed if they are implemented. These attackers can strategically manipulate the data to exploit known vulnerabilities, potentially leading to severe disruptions in traffic management and safety. Typical white-box attacks include GNNS spoofing (Schmidt et al., 2016; Wang et al., 2023a), false data injection attack against navigation systems (Cui et al., 2019), Sybil attack to produce fake CVs (Guette and Ducourthial, 2007), and spoofing attack against TSEP (Wang et al., 2024a). On the other hand, it could be impractical to assume full knowledge of target models and data, which are difficult to obtain in practice. In this case, attacks are not prevalent and thus of low likelihood to initiate due to the difficulty to launch them. For instance, some production-grade data learning systems, including CAV-based obstacle detectors (Sun et al., 2021a) and intelligent RSUs (Wu et al., 2020), typically do not disclose the models and data, making it challenging (if not possible) to obtain the relevant knowledge to launch a white-box attack. In contrast, black-box attacks, which require no detailed insight into the system, are practical to implement. Yet, they also face significant challenges. These attackers typically rely on a trial-and-error approach, including Spoofing attack on LiDAR (Petit et al., 2015; Shin et al., 2017; Yang et al., 2021), attacks on vehicle trajectory planning (Deng et al., 2021), and poisoning attack against traffic prediction (Liu et al., 2022; Zhu et al., 2023). They may struggle to fine-tune their attacks due to limited insights into the system's responses, making it harder to inject malicious data successfully. In summary, assessing the attack likelihood needs to consider the requirement of attacker's knowledge and the practicality such a requirement, and thus could depend on the actual situation where a specific attack is launched.

*Attacker's capability*: Attackers who can manipulate inference datasets have an advantage in evaluating attack model performance at real-time decision time compared to those with training datasets only (Table 5-Table 7). Testing datasets present runtime, real-world scenarios that may not be encountered during training, allowing attackers to assess how well the attack model handles unforeseen situations and adjust the attack to increase the success rate. Spoofing attacks on LiDAR and SLAM, for example, have shown higher success rate, as the vehicle's operation environment could change over time and an attacks against testing dataset could handle the changes well (Petit et al., 2015; Shin et al., 2017; Wang et al., 2021b; Yang et al., 2021). Perturbing feature values is stealthier than flipping labels. Specifically, perturbing feature values involves making small changes to the data itself, such as slightly altering GNNS signals (Schmidt et al., 2016; Wang et al., 2023a) and RSU data (Sumra et al., 2013; Zhao et al., 2022b), which can be more subtle and harder to detect compared to flipping labels. These subtle changes might not immediately raise suspicion, allowing the attack to remain unnoticed for a longer duration. In contrast, flipping labels, which directly altering the data labels (e.g., images from cameras (Lu et al., 2017; Petit et al., 2015)), could lead to immediate and noticeable misclassifications or anomalies, making the attack more evident to detect. Therefore, perturbing feature values is likely associated with a higher success rate.

*Attack strategies*: A successful attack often needs to be application-specific and consider the characteristics and dynamics of the victim system (Table 5-Table 7). There, a bi-level optimization-based attack strategy (e.g., attacks against TSEP (Wang et al., 2024a)) with the ITS solution or system constructed as dynamic system can effectively launch successful attacks. The produced poisoning samples following such strategies are likely optimized for the victim ITS system. Nevertheless, the attack strategy could struggle with accurately capturing system dynamics, which may evolve under various scenarios and suffer from system or random errors. Consequently, the designed data perturbations may become less effective, resulting in medium attack likelihood. On the other hand, for complex systems with unclear objective functions (e.g., attacks against LiDAR (Shin et al., 2017; Yang et al., 2021)), the attack strategy's performance may be compromised due to the involved approximations.

## 5. Discussions of Research Challenges

Our review shows that each of the three primary data sources is threatened by DPAs. The attacks can be categorized following the proposed framework, with risk assessment allowing for evaluating the impacts and likelihood of these attacks. Despite the extensive studies on DPAs against ITS, some research challenges and limitations remain.

### 5.1. Inaccurate and Inefficient Solution to the Attack Model

Bilevel optimization-based attack strategies have demonstrated the potential to significantly enhance attack effectiveness and stealthiness, particularly in ITS applications. In practical implementation, the bilevel optimization-based attack strategy is typically tackled using gradient-based methods (Biggio et al., 2013; Jagielski et al., 2018; Mei and Zhu, 2015; Xiao et al., 2015). The attacker follows an iterative process to generate poisoning data points by perturbing the data in a direction that aligns with the adversarial objective. This direction is determined by computing the gradient of the objective over data perturbations. Specifically, under very restrictive assumptions on the underlying ITS model (i.e., Problem (1-b)), including the smoothness and strong convexity of $S$ and no constraint, the gradient of the upper objective function $O$ over data perturbation $\delta$ can be expressed using the chain rule: $\nabla_\delta O = \nabla_{\theta^*} O(V, M, \theta^*(\delta)) \nabla_\delta \theta^*(\delta)$. By the classical implicit function theorem (Dontchev and Rockafellar, 2009) and optimality condition $\nabla_{\theta^*} S = 0$, $\nabla_\delta \theta^*(\delta)$ can be expressed: $\nabla_\delta \theta^*(\delta) = -\nabla^2_{\delta \theta^*} S \left[ \nabla^2_{\theta^* \theta^*} S \right]^{-1}$, based on which $\nabla_\delta O$ can be derived.

Nevertheless, gradient-based methods face two limitations (Wang et al., 2024a). Firstly, their capability to address constraints, especially inequality constraints that are prevalent in real-world ITS applications, is limited. These constraints encompass physical and domain-specific limitations, shaping the model's solution and interaction with the data to ensure that perturbations are reasonably bounded, thus rendering the attack practical and challenging to detect. The presence of general constraints introduces complexities in analyzing the model's responses to data perturbations: the optimality condition $\nabla_{\theta^*} S = 0$ does not hold and should be replaced by KKT conditions, for which $\nabla_\delta \theta^*(\delta)$ can no longer be explicitly expressed. As shown in Wang et al. (2024), in cases where constraints interact with the solution of the lower-level problem, the computed gradient may not accurately represent the optimal direction for perturbing the data when specific constraints are active at a given data point. Secondly, most existing methods assume that the objective function or constraints are differentiable concerning data changes (Biggio et al., 2013). This assumption, however, does not universally hold. The model solution ($\theta^*(\delta)$) is often continuous but lacks differentiability concerning $\delta$ (Luo et al., 1996), which may lead to the nonexistence of gradient of $\theta^*(\delta)$. Consequently, applying the gradient methods becomes less straightforward.

To address these limitations, Wang et al., (2024) applied the concept of Lipschitz continuity and the semi-derivative based on weaker conditions. As the gradient $\nabla_\delta \theta^*(\delta)$ may not exist, the authors defined the *semi-derivative* of $\theta^*$ over $\delta$, as a counterpart of gradient, to capture how $\theta^*(\delta)$ changes with $\delta$: A function $y$ is said to be semi-differentiable at $\bar{x}$ if it has a first-order approximation at $\bar{x}$ of the form $h(x) = y(\bar{x}) + Dy(x - \bar{x})$ with $Dy$ continuous and positively homogeneous, and the function $Dy$ is called the semi-derivative of $y$ at $\bar{x}$. Following Lipschitz continuity of the solution $\theta^*(\delta)$ over $\delta$, a weaker condition than the differentiability assumption needed for the gradient method, the semi-derivative of $\theta^*(\delta)$ over $\delta$ can be evaluated by using the general implicit function theorem to handle general constraints (Dontchev and Rockafellar, 2009). Then, attackers can select the data perturbation with the largest semi-derivative among all feasible perturbations each time to ensure effective attack.

Despite the advances, the semi-derivative-based method focuses on single-valued mapping and optimization-based learning problems, with limited efforts on set-valued mapping and ML- and DL-based problems. Another challenge is addressing the computational complexity of solving gradients or semi-derivatives, which can be particularly demanding, especially when the number of parameters extends into the millions. Therefore, future research efforts should be directed toward developing solutions to mitigate the complexity inherent in the bi-level poisoning problem (1). One promising direction is to utilize the advances in the field of hyperparameter optimization, which aims to identify an optimal set of hyperparameters associated with a data-learning problem (e.g., the massive structure of DL models). See more discussions in Section 6.1.

5.2. Unrealistic Attack Settings

Many existing DPAs targeting ITS fail to consider realistic settings. While these unrealistic DPAs play a valuable role in stress-testing system robustness under worst-case scenarios, the discrepancy between attack models and real-world ITS intricacies can lead to failure of the proposed attacks in field tests. Many DPAs (e.g., those in white-box settings) may overestimate the attacker's knowledge of the system, such as the logic governing TSC or travelers' responses to fake traffic jams. These assumptions may not align with the real-world complexities and safeguards within ITS environments. Furthermore, it is vital to recognize the dynamic nature of ITS systems, where traffic patterns, infrastructure configurations, and vehicle behaviors continuously evolve. Attackers may find it challenging to sustain and adapt DPAs over extended periods, given these changes and the need to maintain relevance in ever-shifting ITS scenarios.

Additionally, the attackers' capabilities are sometimes exaggerated. For instance, DPAs against infrastructure often unrealistically assume that vulnerabilities within traffic detectors and TSC are easily exploitable and that their values and control parameters can be readily manipulated. Real-world training processes often involve multiple parties, who may identify the coarse perturbations injected to the training data. Existing attacks may also lack justification for the incentive behind an attack, which is equally crucial to make it realistic. In practice, real-world attackers often act with clear motives, such as financial gain, personal benefit, or specific objectives. They could also be bound by legal and ethical constraints, which may deter them from engaging in DPAs. Assessing these incentives and disincentives is fundamental for evaluating the feasibility and likelihood of attacks in practical scenarios.

5.3. Limited Studies on Risk Assessment

Despite the insights summarized in Section 4.5, the assessment framework in our review is largely conceptual, with a motivation to call for more research in this aspect (see Section 6.3). Research quantifying the risk, including impact and likelihood, is under-explored due to several challenges. Firstly, the complexity of ITS components poses a significant challenge. As shown in Figure 3, ITS are composed of interdependent components. Understanding the interdependency among these components requires comprehensive analyses to identify potential attack vectors that attackers might exploit. It is possible that attackers manage to induce broad impacts by leveraging the interdependencies. Another challenge lies in the diverse attack vectors that can be targeted, such as compromising sensor inputs, manipulating communication channels, or tampering with centralized control algorithms. The impacts and likelihood of attacks could vary with different vectors. Yet, identifying all these vectors demands a comprehensive understanding of the system's architecture and the specific vulnerabilities inherent in each component. Moreover, the dynamic nature of ITS adds complexity to risk assessment, as the changing traffic conditions makes it challenging to anticipate potential attack scenarios and their impact across the varying traffic situations.

Furthermore, unlike established attacks in the cybersecurity domain, there might lack historical data specifically related to DPAs in ITS that are primarily in the research and development phase, making it challenging to evaluate the resultant risks based on past occurrences. The conceptual risk assessment framework is subjective due to the lack of support from existing studies and historical data. Sheehan et al. (2019) proposed a Bayesian Network framework for quantifying risk scores of cybersecurity threats in ITS. However, this framework relies on data sources largely irrelevant to ITS, and the learned Bayesian Network may not reflect real-world risks. Specifically, the framework uses the National Vulnerability Database, a large data repository of known software vulnerabilities that aggregates descriptions, references, product names, and Common Vulnerability Scoring System scoring information. Given its focus on general software, this database may have limited utility for assessing DPAs against ITS.

## 6. Future Work

Here we discuss future research needs to address the three challenges discussed earlier (Section 6.1-Section 6.3) and to design more diverse and effective attack models (Section 6.4), and defense solutions (Section 6.5).

### 6.1. Solving the Attack Model

It is imperative to tackle the challenges associated with solving attack models. As noted earlier, one promising approach is applying the Lipschitz analysis and semi-derivative-based methods that were proposed recently relying on weaker conditions (Wang et al., 2024a). Future research needs to extend the Lipschitz analysis method and the calculation of semi-derivatives to set-valued maps, i.e., when the underlying learning model has a set of solutions instead of a single solution. Extensions are also needed for the increasingly common ML- and DL-based ITS applications, whose objective function can be complex and often unknown due to the large number of parameters and the varied model structures. A potential method is to use the "estimator function" to approximate the complex objective function of each layer and apply the Lipschitz analysis methods (Dontchev and Rockafellar, 2009; Wang et al., 2024a).

Another crucial research direction involves finding more scalable DPAs in practical scenarios where solutions could be computationally expensive. To address this challenge, researchers can explore connections between attack models and other research domains formulated within the mathematical framework of bilevel programming. These domains, like hyperparameter optimization, have dedicated significant efforts to address computation-related challenges. Hyperparameter optimization is to determine the optimal combination of hyperparameters that maximizes the performance of an underlying learning algorithm (e.g., the number of layers and neurons in a deep neural network). Identifying an optimal set of hyperparameters is naturally challenging due to the massive scale of the learning algorithm (in terms of both dataset and learning parameters). Utilizing a bilevel framework for formulating DPAs suggests the potential for adapting strategies designed to enhance the optimization of bilevel programs, particularly those associated with hyperparameter optimization tasks. In essence, by conceptualizing poisoning samples as parameters controlled by the attacker, we could explore applying techniques derived from hyperparameter optimization to craft scalable and innovative attacks effectively.

6.2. Considering Realistic Attack Settings

Future research endeavors should evaluate the practicality of DPAs and adopt assumptions that are more realistic. For example, while some models assume that attackers have almost absolute control over the traffic dataset, exploring the viability of DPAs when attackers have limited influence over a small fraction of the training data is equally vital. Besides the datasets, rationalizing the attacker's expertise and resources in gaining access to specific software or hardware facilities is also important. Given most of existing DPAs in ITS focus on white-box settings, future work should explore black- or grey-box settings where attackers' knowledge on victim models or datasets is more realistically reflected. Introducing constraints, such as budget limitations that limit the number of detectors an attacker can manipulate, may provide a more realistic portrayal of the attack scenario. These scenarios embracing less unrealistic assumptions and align better with real-world conditions, which are more valuable for assessing potential threats to actual ITS systems.

Moreover, it is crucial to account for potential defenses when examining attackers' ability to successfully launch stealthy attacks. For instance, established services like Google Maps employ data filters to identify and filter out certain false location data uploaded to their databases. Furthermore, deploying and evaluating these attacks in fields is equally imperative for designing realistic attack scenarios. As previously discussed, real-world ITS systems are characterized by their dynamic nature. Therefore, DPAs should be crafted with this dynamism in mind, as it is conceivable that the effectiveness of poisoning samples may diminish as the system evolves.

6.3. Supporting Risk Assessment

As aforementioned, research on risk assessment of DPA is currently lacking. Based on the challenges identified earlier, several key strategies can be pursued. Firstly, advanced threat modeling and simulation tools should be developed to accurately depict the intricate dependency among ITS components. Attention should be given to scenarios that attacks induce high impacts by leveraging the dependency. Secondly, the risk assessment should encompass a wide array of data poisoning vectors and their potential impacts on systems of all scales. Thirdly, by employing simulations that mirror real-world conditions (e.g., digital-twin ITS testbeds), researchers and practitioners can pay attention to the role of the ITS dynamics, which helps to better understand consequences of different attack scenarios across dynamic traffic environments.

Last but not least, encouraging collaboration between public agencies, private entities, academia, and cybersecurity experts is vital. The collaboration benefits establishing platforms for sharing threat intelligence and best practices and, in particular, encouraging data sharing allow for quantitative, data-driven evaluation of impact and likelihood of attacks (e.g., by setting thresholds for each category).

Moreover, it is critical to advocate developing regulatory frameworks and industry standards specific to ITS cybersecurity, which would offer guidelines for securing ITS components, as well as risk assessment. These data-driven methods and standards are expected to improve the objectivity of the risk assessment that is essential for maintaining fairness and credibility and enhancing the effectiveness of risk management efforts.

6.4. Designing More Diverse and Effective Attack Models

As mentioned earlier, despite the many threats to vehicles, DPAs against pedestrians, cyclists and other road users are rarely studied. Future work could include examinations of how a DPA could affect sensor data or traffic controllers in ways that compromise the safety of these road users, as well as how it could influence the decision-making processes of CAVs when interacting with pedestrians and cyclists. Sensor data and traffic controllers (e.g., for actively avoiding pedestrian/cyclist collision) serving these road users could be different from vehicles and thus present unique challenges in understanding the vulnerabilities (Malik et al., 2020). Future research could support proactive protections of pedestrians, cyclists, and other road users from DPAs.

Another promising research direction is to investigate more effective attack strategies targeting ITS applications, such as the backdoor attack that has demonstrated greater efficacy than other DPAs. As mentioned in Section 4.4, a backdoor attack involves contaminating a model so that a specific trigger embedded in a test input can provoke a chosen model response determined by the adversary. Backdoor attacks exhibit similarities to DPAs discussed earlier but can be *more effective* in specific applications (e.g., CAV's perception systems) (Cinà et al., 2023; Wenger et al., 2021). For instance, attackers may place trigger stickers on traffic signs to deceive a vehicle's camera system, triggering false responses that could compromise vehicle performance (Wang et al., 2021c). Evaluating the effectiveness of backdoor attacks in the context of ITS is an evolving study area. Future research should draw inspiration from adversarial examples from the cybersecurity domain (Cinà et al., 2023) to evaluate the threats of backdoor attacks to the broad ITS applications.

6.5. Designing Defense Solutions

One of the key reasons to study DPAs is to help understand the attacks and develop effective defense methods. Thus, another important future research direction is to develop effective defense methods against DPAs. Numerous defense strategies have been proposed in the literature, primarily aimed at detecting attacks and cleansing both the poisoned data and the underlying model. Given our focus on attack models, here we briefly summarize existing defense methods, their limitations and future directions. Interested readers are referred to the literature review section of the authors' working paper dedicated to defending DPAs in ITS (Wang et al., 2023b).

Most existing detection methods in the transportation community are based on anomaly detection, which detects data attacks by checking abnormal patterns in the training data or incoming inference data. The goal is to identify and eliminate poisoning samples by recognizing the distinctive characteristics of poisoning samples, which often deviate from the pristine training data distribution. The problem is often formulated as an anomaly detection problem and is solved via data-driven or model-based approaches. The *data-driven* approaches rely on previously prepared data to learn a set of patterns or rules, with which the data is determined benign or adversarial (Huang et al., 2021b; van Wyk et al., 2020). The rules could be learned by formulating a supervised or unsupervised learning problem (Luo et al., 2021). The *model-based* detection methods monitor the inconsistency between the incoming data with the expected data that is computed/predicted from a transportation model such as vehicular or traffic dynamic model (Abdollahi Biron et al., 2018; Mousavinejad et al., 2020).

Future works are needed to address the *limitations* in the current defense methods due to the stealthiness of DPAs. Specifically, it is revealed that the anomaly detection model itself could be insecure or poisoned by attacks if the data used for learning the anomaly detector is not guaranteed secure (Shen et al., 2020). Moreover, the system can detect the existence of attacks but may fail to identify the source of attacks (Wang et al., 2023a). Some recent defense methods focus on *adversarial training or certified training* that involve training DL models with both pristine and poisoned data, enhancing model resilience to attacks (Akhtar and Mian, 2018). However, the trained model may suffer from reduced accuracy even as their robustness improves. A promising approach to address these limitations is to *harness secure data from the infrastructure*, enabling the development of "infrastructure-enabled" defense solutions against DPAs. This concept has already been demonstrated success in defending against stealthy GNSS data attacks (Wang et al., 2023a) and some ITS applications such as queue length estimation (Wang et al., 2023b). Exploring infrastructure-enabled defense methods in ITS is emerging, considering that transportation infrastructure is becoming more sophisticated and equipped with sensing and computation capabilities to support advanced transportation applications.

In addition to reactive detection and mitigation, it is promising to design proactive defense strategies. This type of strategy aims to deter attacks by designing the system in a way that is robust to potential attacks (i.e., the "security by design" principle). The field of robust statistics provides tools to limit the impact of a small fraction of adversarial samples. In the emerging field of adversarial machine learning, robust learning algorithms are designed to defend against potential data poisoning attacks, which can be generally grouped into three groups, including data sub-sampling-based estimation, outlier removal methods and trimmed optimization methods (Vorobeychik and Kantarcioglu, 2018). Data sub-sampling-based estimation learns many models using the same learning algorithm, while each model is based on a random sample of the training dataset. Then the model with the smallest training error is selected. Outlier removal methods, as indicated by their name, remove outliers before learning a model from data (Klivans et al., 2009). Trimmed optimization-based methods minimize the empirical risk of the learning model while pruning out a small proportion of data points that incur the largest error (Liu et al., 2017).

Another promising proactive defense strategy could incorporate the Lipschitz analysis, a mathematical concept that gauges how quickly a function or process can change, using the Lipschitz constant to measure the rate of change. When Lipschitz continuity does not hold, a model's solution may change abruptly with a slight perturbation in the data, signifying a non-existent Lipschitz constant. In DL research, studies have been conducted to estimate the Lipschitz constants of models to evaluate their robustness during the model inference (i.e., decision making) stage, which can also be used to assess a model's resilience against DPAs (Scaman and Virmaux, 2019). However, a fundamental challenge remains in estimating the Lipchitz constant during the training stage of a DL model. Furthermore, designing DL models with inherent Lipschitz properties requires more work in the future.

## 7. Conclusion

In conclusion, the increasing reliance of ITS on data exposes them to the emerging threat of DPAs. To the best of our knowledge, this survey paper represents the first comprehensive review that delves into DPAs targeting all primary data sources within the realm of ITS. We have categorized existing DPA models in ITS based on the target data sources and the functions these data support. Furthermore, we have introduced a generic framework for DPAs, focusing on its main components and employing a bilevel framework demonstrating its versatility across a spectrum of attack models. The generic framework enables us to gain valuable insights, particularly regarding risk assessment for these attacks. By critically examining existing attack models, identifying limitations, and highlighting open questions and future research needs in the field of DPAs against ITS, our work aims to raise awareness of the importance of DPA and serve as a vital guideline for a deeper comprehension of the relevant literature. This understanding not only offers insights for risk assessment and the enhancement of defense solutions but also provides a forward-looking perspective on the development of trustworthy ITS, thereby stimulating others to add their thoughts in this important and emerging research field. As ITS evolves, our research endeavors to foster a safer and more resilient future for these critical applications.

**Acknowledgement**

We thank the two anonymous reviewers for their constructive comments that helped significantly improve the original version of the paper. The research is supported by the National Science Foundation (NSF) grant CMMI-2326340. Any opinions, findings, conclusions, recommendations expressed in this paper are those of the authors and do not necessarily reflect the views of NSF.

**Appendix**

A.1. An Example of DPA Attack Components

Figure 6 provides a simple example to help understand the attack components on a Support Vector Machine (SVM) model that uses mobile data to classify vehicles (Sun and Ban, 2013). The two types of data points represent cars and trucks, respectively. The straight line separating the two classes of data stands for the SVM decision boundary (Tax and Laskov, 2003). The attack goal is to shift the SVM decision boundary to maximize the classification errors. A white-box attack is assumed so that the attacker has full knowledge of the SVM model and data. Nevertheless, the attacker is only able to manipulate one vehicle (data point) $x_a$ and cannot change wildly but in a limited region (marked in the dashed circle) to be stealthy (i.e., *attacker's capability*). Given the limit, we identify an effective algorithm, i.e., the *attack strategy*, to reach the poison point $x_a'$ that maximizes the classification errors. As a result, the decision boundary shifts due to the change of support vectors from $(x_b, x_c)$ to $(x_a', x_c)$, and point $x_b$ (and any other samples between the original and poisoned decision boundaries) is misclassified.

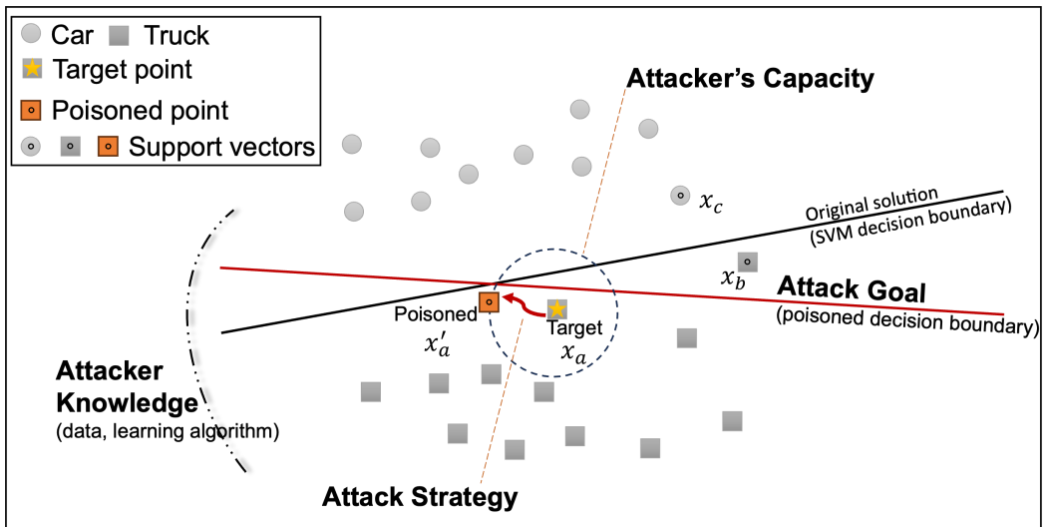

Figure 6. Illustration of DPA against SVM-based vehicle classification

A.2. Risk Assessment of Typical DPAs

Table 5-Table 7 summarize the assessments of impacts and likelihoods of typical DPAs on the three primary ITS data sources. Here, (H, L) means that an attack is of high impact and low likelihood; the same for other categories.

Table 5. Risk assessment of typical DPAs on vehicular data (following Table 1)

| | **Attacks** | **Risk Assessment** |
|---|---|---|
| (Section 3.2.1) Attacks against Vehicle Localization | GNNS spoofing (Schmidt et al., 2016; Wang et al., 2023a) | (H, H) |
| | False data injection attack against SLAM (Wang et al., 2021a) | (M, M) |
| | Attacks on Multi-Sensor Fusion-based localization (Shen et al., 2020) | (H, L) |
| (Section 3.2.2) Attacks against Vehicle Perception | Spoofing attack on camera (Lu et al., 2017; Petit et al., 2015) | (H, M) |
| | Spoofing attack on LiDAR (Petit et al., 2015; Shin et al., 2017; Yang et al., 2021) | (M, L) |
| | Spoofing millimeter radar (Sun et al., 2021b) | (L, L) |
| (Section 3.2.3) Attacks against Vehicle Planning and Control | Attacks on vehicle trajectory planning (Deng et al., 2021) | (H, M) |
| | False data injection attack against navigation systems (Cui et al., 2019) | (H, H) |
| | Compromising AV's driving behavior (Cao et al., 2022) | (H, M) |
| | Manipulating vehicle control (Wang et al., 2021c) | (H, M) |

Table 6. Risk assessment of typical DPAs on V2X communications (following Table 2)

| | **Attacks** | **Risk Assessment** |
|---|---|---|
| (Section 3.3.1) V2V | Sybil attack to produce fake CVs (Guette and Ducourthial, 2007) | (M, M) |
| | False data attack against RSU-supported vehicles (Sumra et al., 2013; Zhao et al., 2022b) | (M, M) |
| | Timing attacks against VANET (Mikki et al., 2013; Rawat et al., 2012; Sumra et al., 2011) | (M, M) |
| (Section 3.3.2) V2I | False data attacks on CV-TSC (Feng et al., 2018; Huang et al., 2021a; Irfan et al., 2022; Yen et al., 2018) | (H, H) |
| (Section 3.3.3) V2N | Spoofing and Sybil attacks against V2N-supported data exchanges (Ristenpart et al., 2009; Salek et al., 2022) | (H, H) |

Table 7. Risk assessment of typical DPAs on infrastructure data (following Table 3)

| | **Attacks** | **Risk Assessment** |
|---|---|---|
| (Section 3.4.1) Traffic management and control | Attacks against fixed-time traffic control systems (Lopez et al., 2020; Perrine et al., 2019) | (M, M) |
| | Data falsify attack against ramp metering control (Ghafouri et al., 2016; Reilly et al., 2016) | (H, M) |
| (Section 3.4.2) Data collection and analysis | DPA at collection (Prigg, 2014) or storage (Cao et al., 2019b; Vivek and Conner, 2022; Wang et al., 2018; Waniek et al., 2021; Zhao et al., 2021) | (H, H) |
| | DPA against TSEP (Wang et al., 2024a), including traffic prediction (Liu et al., 2022; Zhu et al., 2023) | (H, H) |

| (Section 3.4.3) Safety enhancements and charging infrastructure | False information attack against variable message signs (Kelarestaghi et al., 2018) | (H, L) |
|---|---|---|
| | False information attack against RSU data (Hadded et al., 2020) | (H, M) |
| | False data injection attacks on charging infrastructure (Acharya et al., 2020; Guo et al., 2023) | (H, M) |